\documentclass[journal]{IEEEtran}
\ifCLASSINFOpdf
\else
\fi
\usepackage[utf8]{inputenc}
\usepackage{graphicx}
\usepackage{amsmath} 
\usepackage{mathtools} 
\usepackage{amssymb}  
\usepackage{psfrag}
\usepackage{mathrsfs}
\usepackage{xparse}
\usepackage{color}
\usepackage{cases}
\usepackage{pgfplots}
\usepackage{tikz}
\usepackage{pgf-extrect}
\usepackage[authormarkup=none]{changes}
\usetikzlibrary{intersections, positioning,fit,arrows.meta,backgrounds}
\usepackage[final,activate={true,nocompatibility},shrink=10,stretch=10]{microtype}
\usepackage{bm} 

\newtheorem{thm}{Theorem}
\newtheorem{Lemma}{Lemma}
\newtheorem{rem}{Remark}

\newcommand{\col}[1]{\operatorname{col}(#1)}

\renewcommand{\d}{\mathrm{d}}
\newcommand{\e}{\mathrm{e}}

\newcommand{\ie}{i.\,e., }

\newcommand{\del}{\scriptstyle \Delta \displaystyle}

\renewcommand{\t}{^{\top}}

\definechangesauthor[name=Simon,color=blue]{SK} 

\makeatletter
	\def\parsenode[#1]#2\pgf@nil{%
	    \tikzset{label node/.style={#1}}
	    \def\nodetext{#2}
	}
	
	\tikzset{
	    add node at x/.style 2 args={
	        name path global=plot line,
	        /pgfplots/execute at end plot visualization/.append={
	                \begingroup
	                \@ifnextchar[{\parsenode}{\parsenode[]}#2\pgf@nil
	            \path [name path global = position line #1-1]
	                ({axis cs:#1,0}|-{rel axis cs:0,0}) --
	                ({axis cs:#1,0}|-{rel axis cs:0,1});
	            \path [xshift=1pt, name path global = position line #1-2]
	                ({axis cs:#1,0}|-{rel axis cs:0,0}) --
	                ({axis cs:#1,0}|-{rel axis cs:0,1});
	            \path [
	                name intersections={
	                    of={plot line and position line #1-1},
	                    name=left intersection
	                },
	                name intersections={
	                    of={plot line and position line #1-2},
	                    name=right intersection
	                },
	                label node/.append style={pos=1}
	            ] (left intersection-1) -- (right intersection-1)
	            node [label node]{\nodetext};
	            \endgroup
	        }
	    },
	    add node at y/.style 2 args={
	        name path global=plot line,
	        /pgfplots/execute at end plot visualization/.append={
	                \begingroup
	                \@ifnextchar[{\parsenode}{\parsenode[]}#2\pgf@nil
	            \path [name path global = position line #1-1]
	                ({axis cs:0,#1}-|{rel axis cs:0,0}) --
	                ({axis cs:0,#1}-|{rel axis cs:1,1});
	            \path [yshift=1pt, name path global = position line #1-2]
	                ({axis cs:0,#1}-|{rel axis cs:0,0}) --
	                ({axis cs:0,#1}-|{rel axis cs:1,1});
	            \path [
	                name intersections={
	                    of={plot line and position line #1-1},
	                    name=left intersection
	                },
	                name intersections={
	                    of={plot line and position line #1-2},
	                    name=right intersection
	                },
	                label node/.append style={pos=1}
	            ] (left intersection-1) -- (right intersection-1)
	            node [label node] {\nodetext};
	            \endgroup
	        }
	    }
	}
\makeatother

\newcommand*\circled[1]{\tikz[baseline=(char.base)]{
		\node[shape=circle,draw,inner sep=0.5pt] (char) {#1};}}
	
\begin{document}
%
\title{Output Feedback Control of Coupled Linear Parabolic ODE-PDE-ODE Systems }
%
%
%

\author{Joachim~Deutscher,~\IEEEmembership{Member,~IEEE,} and Nicole~Gehring
\thanks{J. Deutscher is with the Institut für Mess-, Regel- und Mikrotechnik, Universit\"at
    Ulm, Albert-Einstein-Allee 41, D-89081~Ulm, Germany.
    (e-mail: joachim.deutscher@uni-ulm.de)\newline
    N. Gehring is with the Institut f\"ur Regelungstechnik und Prozessautomatisierung, Universit\"at Linz,  Altenberger Stra{\ss}e 69, 4040~Linz, Austria. (e-mail: nicole.gehring@jku.at)}
}

%
%

\markboth{IEEE TRANSACTIONS ON AUTOMATIC CONTROL,~Vol.~XX, No.~Y, JULY~2020}%
{Shell \MakeLowercase{\textit{et al.}}: Bare Demo of IEEEtran.cls for Journals}
%



\maketitle

\begin{abstract}
This paper deals with the backstepping design of observer-based compensators for parabolic ODE-PDE-ODE systems. The latter consist of $\pmb{n}$ coupled parabolic PDEs with distinct diffusion coefficients and spatially-varying coefficients, that are bidirectionally coupled to ODEs at both boundaries. The actuation and sensing appears through these ODEs resulting in a challenging control problem. For this setup a systematic backstepping approach is proposed, in order to determine a state feedback controller and an observer. In particular, the state feedback loop and the observer error dynamics are mapped into stable ODE-PDE-ODE cascades by making use of a sequence of transformations. With this, the design can be traced back to the solution of kernel equations already found in the literature as well as initial and boundary value problems, that can be solved numerically. Exponential stability of the closed-loop system is verified, wherein the decay rate can be directly specified in the design. The results of the paper are illustrated by the output feedback control of an unstable ODE-PDE-ODE system with two coupled parabolic PDEs.


\end{abstract}

\begin{IEEEkeywords}
Distributed-parameter systems, parabolic systems, ODE-PDE-ODE systems, backstepping, boundary control, observers. 
\end{IEEEkeywords}

\section{Introduction}
\subsection{Motivation and Literature Review}
Recent advances of the \emph{backstepping approach} (see, e.\:g., \cite{Kr08} for an overview) focused on the control of coupled PDEs. In the parabolic case, one can find backstepping methods for this system class with constant coefficients in \cite{Ba15a,Orl17}, while the general case of spatially-varying coefficients is dealt with in \cite{Vaz16a,Ca20,Deu17}. These results provide systematic solutions for the control of a variety of applications. Typical examples appear in chemical and biochemical engineering problems (see \cite{Ja14,At74}).

More complex models also take finite-dimensional actuator and sensor dynamics into account. This naturally leads to parabolic \emph{ODE-PDE-ODE cascades}. It should be noted that this setup is fundamentally different from the backstepping approach for PDE-ODE cascades, in which the actuator or sensor is an infinite-dimensional system and the plant is finite-dimensional (see \cite{Kr09} and the references therein). In particular, the modelling of actuators and sensors with ODEs results in a much more challenging control problem, because only lumped actuation and sensing is possible. A more general class of systems are \emph{coupled ODE-PDE-ODE systems}, where the actuating and sensing ODEs are bidirectionally coupled to the PDEs. Such a setup results for coupled PDEs with dynamic boundary conditions. The latter occur, for example, in the modelling of solid-gas interactions of chemical reactions at the boundaries (see, e.\:g., \cite{Ta11}). For a single parabolic PDE, which is bidirectionally coupled to an ODE at the unactuated boundary, the result \cite{Ta11} provides a backstepping solution for the state feedback design assuming an actuating PDE, and for the observer design, considering a boundary sensing at the PDE. 

So far, only few results exist for the backstepping control of parabolic ODE-PDE-ODE systems in the literature. A first approach is presented in  \cite{Wa19} for a single parabolic PDE. Therein, the backstepping state feedback control of a heat equation cascaded with an actuating ODE at one boundary and a bidirectionally coupled ODE at the other boundary is presented. In order to implement this controller, a backstepping observer for a boundary measurement of the heat equation is designed. For hyperbolic systems more results are available in the literature. A first solution for a hyperbolic ODE-PDE-ODE cascade, in which the PDE is a single transport equation, can be found in \cite{An18}. The case of coupled hyperbolic ODE-PDE-ODE systems is dealt with in  \cite{Sab17,Wa18}, where the PDEs describe a $2 \times 2$ hyperbolic system. Recently, \cite{Deu18a} considers a general setup, where the PDEs represent a general heterodirectional hyperbolic system. However, in contrast to \cite{Sab17,Wa18}, the design yields a target system that is a stable ODE-PDE-ODE cascade. This significantly facilitates the stability analysis and allows to specify the finite-dimensional part of the closed-loop dynamics by a simple eigenvalue assignment for the involved ODEs.

In order to determine backstepping controllers for coupled ODE-PDE-ODE systems, two possible methods can be found in the literature. The first one is the \emph{one step approach}, in which the plant is mapped into a final target system in one step (see \cite{Wa19} for the parabolic case and \cite{Sab17,Wa18} for hyperbolic systems). The corresponding backstepping transformation is obtained from the solution of the resulting kernel equations. However, the coupled structure of the ODE-PDE-ODE system leads to kernel equations, which consist of PDEs coupled to ODEs. Hence, it may be hard to determine the corresponding kernel. Furthermore, the final target system is still a coupled ODE-PDE-ODE system, which impedes the corresponding stability analysis.

The second method is the \emph{successive backstepping approach} first proposed in \cite{Deu18a} for hyperbolic systems, where the system is mapped into a final target system in multiple steps.  Therein, only the kernel equations consisting of a system of PDEs have to be solved, in order to simplify the PDE subsystem. For this, systematic procedures can already be found in the literature. As the structure of the intermediate target systems becomes increasingly simple from step to step, also the boundary value problems (BVPs) and initial value problems (IVPs) to be solved for calculating the transformations take a simple form. In some sense, the coupled kernel equations resulting from the one step approach are decomposed into several easily solvable equations. With this, it is possible to determine backstepping controllers for very general setups in an easy way. This was recently demonstrated for general hyperbolic coupled ODE-PDE-ODE systems in \cite{Deu18a}. The choice of the transformations for the successive backstepping is guided by the coupling structure of the ODE-PDE-ODE system. In particular, firstly, the PDE is mapped into a simpler target system. This significantly facilitates the elimination of couplings between the ODEs and PDEs at both boundaries by the subsequent transformations. As a result, a stable ODE-PDE-ODE cascade is obtained as the final target system. The lumped actuation and sensing is taken into account by mapping the actuating and sensing ODEs into special forms. This allows a systematic derivation of the related decoupling transformations.

\subsection{Contributions}
In this paper, it is shown that the successive backstepping approach also provides a systematic solution of the output feedback stabilization problem for a very general class of coupled parabolic ODE-PDE-ODE systems. The PDE subsystem consists of $n$ in-domain coupled parabolic PDEs with mutually different diffusion coefficients, spatially-varying parameters and Robin boundary conditions. This system is bidirectionally coupled to linear ODEs at both boundaries. The actuating ODE at $z = 1$ is subject to the input and the anti-collocated measurement only depends on the state of the sensing ODE at $z = 0$. This general setup has, so far, not been investigated even for hyperbolic systems and leads to new significant challenges in the design, for which systematic solutions are provided in the paper.

The first part of the paper deals with the backstepping design of a state feedback controller. The latter is derived from mapping the initially bidirectionally coupled and possibly unstable ODE-PDE-ODE system into a stable ODE-PDE-ODE cascade. In comparison to the backstepping control of coupled hyperbolic ODE-PDE-ODE systems in \cite{Deu18a}, a different sequence of transformations is utilized for the state feedback design to map the plant into a stable ODE-PDE-ODE cascade. The result in \cite{Deu18a} requires the solution of the usual kernel equations and a set of coupled Volterra integral transformations of the second kind, which  are successively derived from explicitly solving an IVP for an ODE and a BVP for a PDE with the method of characteristics. This approach, however, is no longer feasible for parabolic PDEs due to the second order derivatives in the IVP and BVP. Therefore, a new approach is proposed in the paper, which numerically solves only the usual kernel equations found in \cite{Deu17} and an IVP. This is crucial in order to obtain a systematic design procedure for the backstepping control of coupled parabolic ODE-PDE-ODE systems. This part of the paper is concluded by verifying the exponential stability of the resulting closed-loop system. The corresponding stability result demonstrates that the closed-loop decay rate can be directly specified by an eigenvalue assignment for the ODEs and choosing a suitable target system for the PDE subsystem. 

In order to obtain an output feedback controller, a state observer is designed in the second part of the paper. This problem is considered for an anti-collocated setup by making use of the successive backstepping approach. Consequently, the observer design for coupled hyperbolic ODE-PDE-ODE systems in \cite{Deu18a} is extended, where only a collocated boundary measurement of the PDE was utilized. In particular, the much more challenging problem of the observer design subject to an ODE measurement is considered in the paper. This requires to introduce a non-obvious state space representation of the sensing ODE. The latter can be seen as a \emph{dual Byrnes-Isidori normal form} (see, e.\:g., \cite[Ch. 5.1]{Is95} for the classical Byrnes-Isidori normal form), which was not introduced so far for finite-dimensional systems. By proposing a suitable output injection structure for the corresponding backstepping observer, it is shown that this method leads to a systematic design procedure. More precisely, the design can be traced back to solving the observer kernel equations presented in \cite{Deu18} as well as IVPs and BVPs. Exponential stability is shown for the resulting observer error dynamics, in which the observer error decay rate can be directly specified in the design. 

Finally, it should be remarked that the solvability analysis of the BVPs appearing in the state feedback and observer design is impeded for parabolic systems, because the fundamental matrices utilized in the proofs can no longer be determined explicitly. This is an essential difference to the hyperbolic case considered in \cite{Deu18a}, where these matrices can be obtained in closed form. In order to solve this problem, a new duality between the BVPs for the eigenvalue problem w.r.t. the spatial differential operator and for the decoupling transformations is unveiled. With this, a reciprocity relation can be established for these fundamental matrices allowing to derive solvability conditions for the BVPs in questions.

By utilizing the state estimates of the observer in the state feedback, an observer-based compensator is obtained. For this setup, the validity of the separation principle is proved. Hence, the stability margin of the closed-loop dynamics can be specified in the state feedback and observer design. This results in a systematic backstepping method to determine observer-based compensators for a large class of parabolic ODE-PDE-ODE systems.

\subsection{Organization}
The next section presents a formulation of the considered control problem. Afterwards, the state feedback design is presented in Section \ref{sec:sfeed}. Then, Section \ref{sec:obs} considers the determination of an observer for an anti-collocated setup. By combining the state feedback with the observer, an observer-based compensator results, which is investigated in Section \ref{sec:obscomp}.
The output feedback control of an unstable parabolic ODE-PDE-ODE system with two coupled PDEs in Section \ref{sec:ex} illustrates the results of the paper. The effectiveness of the new controller design procedure is confirmed in simulations.

\emph{Notations}: In the paper, Lagrange's notation for the partial derivative, \ie $f_z = \partial_z f$, and  Leibniz's notation for the ordinary derivative, \ie  $(\,\cdot\,){'} = \frac{\d}{\d z}(\,\cdot\,)$, are sometimes applied. For convenience, the notations $\partial_zf(z,z) = \partial_zf(z,\zeta)|_{\zeta=z}$ and $\partial_{\zeta}f(z,z) = \partial_{\zeta}f(z,\zeta)|_{\zeta=z}$ are introduced.
 
\section{Problem formulation}\label{sec:probform}
Consider the \emph{coupled linear parabolic ODE-PDE-ODE system}
\begin{equation*}
u \to \boxed{\Sigma_{n_1}(w_1)} \underset{z=1}{\Leftrightarrow} \boxed{\Sigma_{\infty}(x)} \underset{z=0}{\Leftrightarrow}\boxed{\Sigma_{n_0}(w_0)} \to y
\end{equation*}
described by
\begin{subequations}\label{drs}
\begin{align}
\partial_tx(z,t) &= \Lambda(z)\partial_z^2x(z,t)  +  A(z)x(z,t)\label{pdes}\\
\partial_zx(0,t) &= Q_0x(0,t) + C_0w_0(t), && t > 0\label{bc1}\\
\partial_zx(1,t) &= Q_1x(1,t) + C_1w_1(t),&& t > 0\label{bc2}\\
    \dot{w}_0(t) &= F_0w_0(t) + B_0x(0,t), &&  t > 0\label{plantode}\\
    \dot{w}_1(t) &= F_1w_1(t) + B_1x(1,t) + Bu(t), &&  t > 0\label{plantode2}\\
            y(t) &= Cw_0(t), &&  t \geq 0,\label{meas}
\end{align}
\end{subequations}
with the distributed state $x(z,t) \in \mathbb{R}^{n}$ and \eqref{pdes} defined on $(z,t) \in (0,1) \times \mathbb{R}^+$. The matrix $\Lambda(z) \in \mathbb{R}^{n \times n}$ in \eqref{pdes} is the diagonal matrix
\begin{equation}\label{Lambdamat}
\Lambda(z) = \operatorname{diag}(\lambda_1(z),\ldots,\lambda_n(z))
\end{equation}
with $\lambda_i \in C^{2}[0,1]$, $i = 1,2,\ldots,n$, satisfying $\lambda_1(z) > \ldots > \lambda_n(z) > 0$, and $A \in (C^1[0,1])^{n \times n}$ is assumed. Furthermore, decoupled Robin boundary conditions (BCs) are considered at both boundaries giving rise to the diagonal matrices
\begin{equation}\label{P1def}
Q_i = \operatorname{diag}(q_i^1,\ldots,q_i^n), \quad i = 0,1.
\end{equation}
The \emph{sensing ODE} \eqref{plantode} has the state $w_0(t) \in \mathbb{R}^{n_0}$, while the state $w_1(t) \in \mathbb{R}^{n_1}$ describes the \emph{actuating ODE} \eqref{plantode2}. The input is $u(t) \in \mathbb{R}^n$ and the anti-collocated output $y(t) \in \mathbb{R}^n$ is available for measurement. The initial condition (IC) of the PDE subsystem is $x(z,0) = x_0(z) \in \mathbb{R}^n$, and the ODE subsystems have the ICs $w_0(0) = w_{0,0} \in \mathbb{R}^{n_0}$, $w_1(0) = w_{1,0} \in \mathbb{R}^{n_1}$.

\begin{rem}
In the general setup of the paper, the ODEs \eqref{plantode} and \eqref{plantode2} represent \emph{dynamic BCs} if the modelling of the plant yields ODEs interacting with the PDEs at the boundaries. Another important situation is the consideration of \emph{finite-dimensional sensor dynamics} and \emph{finite-dimensional actuator dynamics}.	\hfill $\triangleleft$
\end{rem}

\begin{rem}
The class \eqref{drs} of systems typically appears in the modelling of chemical reactors with actuator and sensor dynamics (see, e.\:g., \cite{Ra81,Je82}). Furthermore, the case in which some of the coefficients $\lambda_i$ in \eqref{Lambdamat} are equal can also be considered by a simple extension of the results in \cite{Deu17}. If, in addition, advection is present in the system. i.\:e., the PDE has the more general form $\partial_tx(z,t) = \partial_z(\Lambda(z)\partial_z x(z,t))  + \Phi(z)\partial_zx(z,t) + A(z)x(z,t)$
with
\begin{equation}\label{phicondef}
	\Phi(z) = \operatorname{diag}(\Phi_1(z),\ldots,\Phi_n(z))
\end{equation}
and $\Phi \in (C^1[0,1])^{n \times n}$, then the boundedly invertible \emph{Hopf-Cole-type state transformation}
\begin{equation}\label{hctraf}
\check{x}(z,t) = \exp\left(\tfrac{1}{2}\!\int_0^z\Lambda^{-1}(\zeta)(\Lambda'(\zeta)+\Phi(\zeta))\d \zeta\right)x(z,t)
\end{equation}
can be utilized to trace this PDE back to \eqref{pdes}. This leaves the structure of the BCs unchanged. \hfill $\triangleleft$
\end{rem}

This paper is concerned with the \emph{backstepping design} of an observer-based compensator for \eqref{drs}, in order to ensure exponential stability of the resulting closed-loop system. For this, the following assumptions are imposed:
\begin{enumerate}
	\item  $(F_0,B_0)$ is controllable,\label{contrass}
	\item  $(C_1,F_1)$ is observable,\label{obsass}
	\item  $\det (C_1B) \neq 0$ and \label{deouplu}
    \item  $\det (CB_0) \neq 0$.	\label{rone}
\end{enumerate}
The Assumptions (A\ref{contrass}), (A\ref{obsass}) are sufficient conditions for the stabilization of the state feedback loop and of the observer error dynamics. In particular, these assumptions are required to ensure the stabilizability of the ODE systems $(F_0,B_0)$ by state feedback and of $(C_1,F_1)$ by output injection. It should be noted that Assumptions (A\ref{contrass}), (A\ref{obsass}) can be relaxed to stabilizability and detectability with additional conditions for the uncontrollable and unobservable modes, which directly follows from the results of the paper. Furthermore, the Assumptions (A\ref{deouplu}), (A\ref{rone}) are required in order to map the $w_1$- and $w_0$-systems into their decoupling forms. The latter are needed to determine a sequence of transformations for mapping the state feedback loop and the observer error dynamics into stable ODE-PDE-ODE cascades. Note that the last two assumptions imply $n_1 \geq n$, $n_0 \geq n$ and $\operatorname{rk}C_1 = \operatorname{rk}B = \operatorname{rk}C = \operatorname{rk}B_0 = n$.

\begin{rem}
It should be noted that it is possible to relax Assumptions (A\ref{deouplu}) and (A\ref{rone}). More precisely, these conditions can be satisfied by making use of dynamic input and output extensions. To this end, the results in \cite{Mo70} may be the point of departure to formulate the presented results for a more general setup. Furthermore, Assumption (A\ref{deouplu}) implies that the system $(C_1,F_1,B)$ has vector relative degree one. There are no obstacles to extend the results of the paper with more involved technical developments to a higher relative degree. Then, however, also higher order derivatives may appear in the controller. A similar result also holds for the observer design. \phantom{leer}	\hfill $\triangleleft$
\end{rem}

\section{State feedback design}\label{sec:sfeed}
\begin{figure}[t]
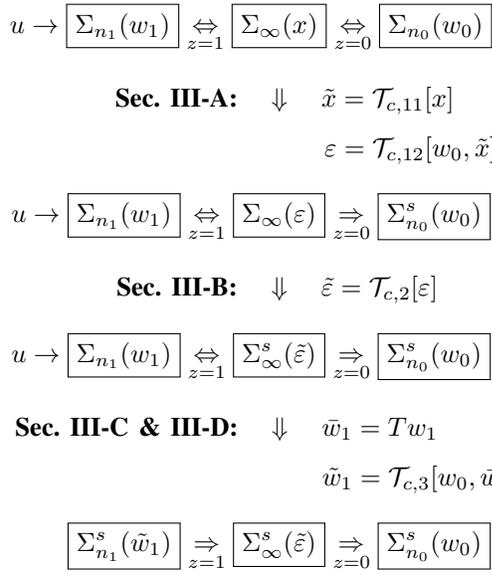

\begin{equation*}
u \to \boxed{\Sigma_{n_1}(w_1)} \underset{z=1}{\Leftrightarrow} \boxed{\Sigma_{\infty}(x)} \underset{z=0}{\Leftrightarrow}\boxed{\Sigma_{n_0}(w_0)}
\end{equation*}
\vspace{-0.1cm}
\begin{equation*}
\hspace{0.9cm}\textbf{Sec. \ref{sec:sf1}:} \quad \Downarrow \quad \tilde{x} = \mathcal{T}_{c,11}[x]
\end{equation*}
\begin{equation*}
\hspace{4.25cm} \varepsilon = \mathcal{T}_{c,12}[w_0,\tilde{x}]
\end{equation*}
\vspace{-0.2cm}
\begin{equation*}
u \to \boxed{\Sigma_{n_1}(w_1)} \underset{z=1}{\Leftrightarrow} \boxed{\Sigma_{\infty}(\varepsilon)} \underset{z=0}{\Rightarrow}\boxed{\Sigma^s_{n_0}(w_0)}
\end{equation*}
\vspace{-0.1cm}
\begin{equation*}
\hspace{0.7cm}\textbf{Sec. \ref{sec:sf2}:} \quad \Downarrow \quad 
 \tilde{\varepsilon} = \mathcal{T}_{c,2}[\varepsilon]
\end{equation*}
\vspace{-0.2cm}
\begin{equation*}
u \to \boxed{\Sigma_{n_1}(w_1)} \underset{z=1}{\Leftrightarrow} \boxed{\Sigma^s_{\infty}(\tilde{\varepsilon})} \underset{z=0}{\Rightarrow}\boxed{\Sigma^s_{n_0}(w_0)}
\end{equation*}
\vspace{-0.1cm}
\begin{equation*}
\hspace{-0.65cm}\textbf{Sec. \ref{sec:decouplw1} \& \ref{sec:sffinal}:} \quad \Downarrow \quad  \bar{w}_1 = Tw_1
\end{equation*}
\begin{equation*}
\hspace{4.82cm}\tilde{w}_1 = \mathcal{T}_{c,3}[w_0,\bar{w}_1,\tilde{\varepsilon}] 
\end{equation*}

\vspace{-0.2cm}
\begin{equation*}
\;\phantom{u \to }\boxed{\Sigma^s_{n_1}(\tilde{w}_1)} \underset{z=1}{\Rightarrow} \boxed{\Sigma^s_{\infty}(\tilde{\varepsilon})} \underset{z=0}{\Rightarrow}\boxed{\Sigma^s_{n_0}(w_0)}
\end{equation*}
\caption{Sequence of transformations mapping the state feedback loop into a stable ODE-PDE-ODE cascade. The $\Sigma^s$-systems are stabilized.}\label{fig:tseq}
\end{figure}

In this section, the \emph{state feedback controller}
is determined. For this, the closed-loop system is mapped into the stable \emph{ODE-PDE-ODE cascade}
\begin{subequations}\label{drsc}
	\begin{align}
	\dot{\tilde{w}}_1(t) &= \tilde{F}_1\tilde{w}_1(t) \label{plantode2c}\\
	\partial_t\tilde{\varepsilon}(z,t) &= \Lambda(z)\partial_z^2\tilde{\varepsilon}(z,t) - \mu_c\tilde{\varepsilon}(z,t) - \tilde{A}_0(z)\tilde{\varepsilon}(0,t)\label{pdesc}\\
	\partial_z\tilde{\varepsilon}(0,t) &= 0\label{bc1c}\\
	\partial_z\tilde{\varepsilon}(1,t) &= D\tilde{w}_1(t)\label{bc2c}\\
	\dot{w}_0(t) &= \tilde{F}_0w_0(t) + B_0\tilde{\varepsilon}(0,t)\label{plantodec}		
	\end{align}
\end{subequations}
with
\begin{subequations}
\begin{align}
 \tilde{F}_1 &= \begin{bmatrix}
  - K_{11}    &  - K_{12}\\
  \tilde{F}_{21} & F_{22}
 \end{bmatrix}\label{F1til}\\
 \tilde{A}_{0}(z) &= \begin{bmatrix} 0     & \ldots &  \ldots           & 0\\
 \tilde{a}_{21}(z) & \ddots       & \ddots             & \vdots\\
 \vdots    & \ddots &  \ddots     &  \vdots \\
 \tilde{a}_{n1}(z) & \ldots & \tilde{a}_{n\,n-1}(z) & 0
 \end{bmatrix}\label{a01def}\\
 D &= \begin{bmatrix}
  I_{n} & 0
 \end{bmatrix}\\
 \tilde{F}_0 &= F_0-B_0K_0\label{F0tildeg}
\end{align}	
\end{subequations}
and the matrices in \eqref{F1til} being introduced in Section \ref{sec:decouplw1}.
Therein, the feedback gains $K_{11}$ and $K_{22}$ ensure stability of the $\tilde{w}_1$-system, i.\:e., that $\tilde{F}_1$ in \eqref{F1til} is a Hurwitz matrix. The strictly lower triangular structure of $\tilde{A}_{0}(z)$ in \eqref{a01def} implies that the  distributed-parameter system (DPS) \eqref{pdesc}--\eqref{bc2c} is a cascade of stable parabolic PDEs for suitable $\mu_c$. Finally, the feedback gain $K_0$ is determined to stabilize the $w_0$-system giving rise to a stable ODE-PDE-ODE cascade \eqref{drsc}. In what follows, the sequence of transformations depicted in Figure \ref{fig:tseq} is presented that map \eqref{drs} into \eqref{drsc}, from which the state feedback controller is determined. The section is concluded by the stability proof for \eqref{drsc} and the related closed-loop stability in the original coordinates.

\subsection{Decoupling of the $\Sigma_{\infty}$-System and Stabilization of the\newline $\Sigma_{n_0}$-System}\label{sec:sf1}
Consider the \emph{backstepping transformation}
\begin{subequations}\label{decoupbacktrafo}
\begin{equation}
 \tilde{x}(z,t) = \mathcal{T}_{c,11}[x(t)](z) = x(z,t)  - \int_0^zK(z,\zeta)x(\zeta,t)\d\zeta\label{Tdef}
\end{equation}
and the \emph{decoupling transformation}
\begin{equation}
 \varepsilon(z,t) = \mathcal{T}_{c,12}[w_0(t),\tilde{x}(t)](z) =  \tilde{x}(z,t) - N(z)w_0(t).\label{etraf}
\end{equation}	 
\end{subequations}
Therein, $K(z,\zeta) \in \mathbb{R}^{n \times n}$, $N(z) \in \mathbb{R}^{n \times n_0}$ are the kernel and the transformation matrix to be determined, while the inverse of \eqref{Tdef} is the backstepping transformation
\begin{equation}\label{Tdefinv}
	x(z,t) = \mathcal{T}^{-1}_{c,11}[\tilde{x}(t)](z) = \tilde{x}(z,t) + \int_0^zK_I(z,\zeta)\tilde{x}(\zeta,t)\d\zeta
\end{equation}
with $K_I(z,\zeta) \in \mathbb{R}^{n \times n}$.
\begin{rem}\label{rem:recrel}
It should be mentioned that the kernel $K_I(z,\zeta)$ can be determined from $K(z,\zeta)$ by making use of the \emph{reciprocity relation}
\begin{equation}
 K_I(z,\zeta) = K(z,\zeta) + \int_{\zeta}^{z}K(z,\bar{\zeta})K_I(\bar{\zeta},\zeta)\d\bar{\zeta},
\end{equation}
following from substituting \eqref{Tdefinv} in \eqref{Tdef} and utilizing a change of the integration order. This is a Volterra integral equation of the second kind, which can be solved with the method of successive approximations. The same results also holds for all other backstepping transformations in the sequel.
\hfill $\triangleleft$
\end{rem}
Then, \eqref{etraf} can be solved for $x(z,t)$ to obtain
\begin{equation}\label{invtraf}
	x(z,t) = \mathcal{T}_{c,11}^{-1}[\varepsilon(t)](z) + N_I(z)w_0(t),
\end{equation}
in which $N_I(z) = \mathcal{T}^{-1}_{c,11}[N](z)$ is introduced in the following. The backstepping transformation \eqref{Tdef} is introduced to remove the coupling term $A(z)x(z,t)$ in the PDE \eqref{pdes}, while the decoupling transformation \eqref{etraf} eliminates the coupling in the BC \eqref{bc1} to the $w_0$-system (see Fig. \ref{fig:tseq}).

Hence, the transformations \eqref{decoupbacktrafo} map \eqref{drs} into the \emph{first intermediate target system}
\begin{subequations}\label{drsc1}
	\begin{align}
	\partial_t\varepsilon(z,t) &= \Lambda(z)\partial_z^2\varepsilon(z,t) - \mu_c\varepsilon(z,t) - H_0(z)\varepsilon(0,t)\label{pdesc1}\\
	\partial_z\varepsilon(0,t) &= 0\label{bc1c1}\\
	\partial_z\varepsilon(1,t) &= \tilde{Q}_1\varepsilon(1,t) + \textstyle\int_0^1K_1(\zeta)\varepsilon(\zeta,t)\d \zeta + \tilde{C}_1w_0(t)\nonumber\\
	& \quad  + C_1w_1(t)\label{bc2c1}\\
	\dot{w}_0(t) &= \tilde{F}_0w_0(t) + B_0\varepsilon(0,t)\label{plantodec2}\\
	\dot{w}_1(t) &= F_1w_1(t)  + \tilde{B}_0w_0(t) + Bu(t) \nonumber \\
	& \quad + B_1\varepsilon(1,t)  + \textstyle\int_0^1B_1K_I(1,\zeta)\varepsilon(\zeta,t)\d \zeta,\label{w1sys1}
	\end{align}
\end{subequations}
in which
\begin{subequations}
	\begin{align}
	 H_0(z) &= A_0(z) + N(z)B_0\label{H0def}\\
	 A_{0}(z) &= \begin{bmatrix} 0     & \ldots &  \ldots           & 0\\
	 a_{21}(z) & \ddots       & \ddots             & \vdots\\
	 \vdots    & \ddots &  \ddots     &  \vdots \\
	 a_{n1}(z) & \ldots & a_{n\,n-1}(z) & 0
	 \end{bmatrix}\label{a01defuntil}\\
	 \tilde{Q}_1 &= Q_1 - K(1,1)\\
	 \bar{K}_1(\zeta) &=  - K_z(1,\zeta) - \int_{\zeta}^1K_z(1,\bar{\zeta})K_I(\bar{\zeta},\zeta)\d\bar{\zeta}\label{K1bardef}\\
	 K_1(\zeta) &= \tilde{Q}_1K_I(1,\zeta) + \bar{K}_1(\zeta)\label{K1def}\\
	  \tilde{C}_1 &= - N'(1) + \int_0^1\bar{K}_1(\zeta)N(\zeta)\d\zeta + \tilde{Q}_1N_I(1)\\
	  \tilde{B}_0 &= B_1N_I(1)\label{lasteqimed}.	  
	\end{align}
\end{subequations}
Therein, \eqref{K1bardef}--\eqref{lasteqimed} are obtained by making use of \eqref{invtraf} and a straightforward calculation including a change of the integration order to represent BC \eqref{bc2c1} and the $w_1$-system \eqref{w1sys1} in the new coordinates. 
The matrix $K_0$ in \eqref{plantodec2} has to ensure that $\tilde{F}_0 = F_0-B_0K_0$ is Hurwitz, which is always possible in view of Assumption (A\ref{contrass}). Hence, the $w_0$-system \eqref{plantodec2} is stabilized in this step.

Differentiate \eqref{etraf} w.r.t. time and insert \eqref{drs} as well as \eqref{drsc1}, by an integration by parts and application of the Leibniz differentiation rule, it can be shown that the kernel has to satisfy the \emph{kernel equations}%
\begin{subequations}\label{keq}
	\begin{align}
	& \Lambda(z) K_{zz}(z,\zeta) - (K(z,\zeta)\Lambda(\zeta))_{\zeta\zeta}  
	= K(z,\zeta)(A(\zeta) + \mu_cI)\label{kpde}\\
	& K_{\zeta}(z,0)\Lambda(0) + K(z,0)(\Lambda'(0) - \Lambda(0)Q_0) = A_0(z)\label{kBC1}\\
	& \Lambda(z) K'(z,z) + \Lambda(z)K_z(z,z) + K_{\zeta}(z,z)\Lambda(z) \nonumber\\
	&\quad  + K(z,z)\Lambda'(z)  = -(A(z) + \mu_cI)\label{kBC2}\\
	& K(z,z)\Lambda(z) - \Lambda(z)K(z,z) = 0\label{kBC3}\\  
	& K(0,0) = Q_0,\label{kIC}
	\end{align}
\end{subequations}
in which \eqref{kpde} is defined on $0 < \zeta < z < 1$.  It is verified in \cite{Deu17} that \eqref{keq} has a piecewise $C^2$-solution. With it, the elements $a_{ij}(z)$, $i > j$, in \eqref{a01defuntil} are determined. Furthermore, it follows that $N(z)$ has to solve the initial value problem (IVP)
\begin{subequations}\label{ivp:sf}
	 \begin{align}
	  &\Lambda(z)N''(z) - \mu_cN(z) - N(z)\tilde{F}_0\nonumber\\
	  &\qquad = -A_0(z)K_0 - K(z,0)\Lambda(0)C_0, \quad z \in (0,1]\\
	  &N(0) = -K_0\\
	  &N'(0) = C_0
	 \end{align}
\end{subequations}
(see \eqref{F0tildeg}). The solvability of \eqref{ivp:sf} is presented in the next lemma.
\begin{Lemma}\label{lem:sflemdecoupl}
The IVP \eqref{ivp:sf} has a unique solution, with the elements of $N$ piecewise $C^2$-functions. 
\end{Lemma}

\begin{IEEEproof}
Introduce $n_{i}\t(z) = e_i\t N(z)$, $h_i\t(z) = e_i\t\big(-A_0(z)\linebreak \cdot K_0 - K(z,0)\Lambda(0)C_0\big)$, $g_{1,i}\t = -e_i\t K_0$ and $g_{2,i}\t = e_i\t C_0$ for $i = 1,2,\ldots,n$, in which $e_i \in \mathbb{R}^n$ is the unit vector. Then, premultiplying \eqref{ivp:sf} by $e_i\t$, $i = 1,2,\ldots,n$, gives
\begin{subequations}\label{ivp:sfa}
		\begin{align}
		\d_z^2n_i\t(z) &= \tfrac{1}{\lambda_i(z)}\big(n_i\t(z)(\mu_cI + \tilde{F}_0) + h_i\t(z)\big)\label{app:ODEdecoupl}\\
		n_i\t(0) &= g_{1,i}\t \\
		\d_zn_i\t(0) &= g_{2,i}\t, 
		\end{align}
\end{subequations}
where \eqref{app:ODEdecoupl} is defined on $z \in (0,1)$. With this, the proof of Lemma \ref{lem:sflemdecoupl} follows from standard results for ODEs.
\end{IEEEproof}

\begin{rem}
For a constant matrix $\Lambda(z) = const.$ in \eqref{pdes}, the solution of \eqref{ivp:sf} can be determined in closed-form by utilizing the matrix exponential. Since the same holds true for all other subsequent IVPs and boundary value problems (BVPs), only the kernel equations of the backstepping transformations have to be solved numerically in this case.  \hfill $\triangleleft$
\end{rem}

\subsection{Stabilization of the $\Sigma_{\infty}$-System}\label{sec:sf2}
Unfortunately, the matrix $H_0(\zeta)$ in \eqref{H0def} has not the form of $\tilde{A}_0(z)$ in \eqref{a01def}, i.\:e., it is not strictly lower triangular. Therefore, the PDEs in \eqref{pdesc1} are still bidirectionally coupled so that the additional \emph{backstepping transformation}
\begin{equation}\label{trafc2}
 \tilde{\varepsilon}(z,t) = \mathcal{T}_{c,2}[\varepsilon(t)](z) = \varepsilon(z,t) - \int_0^zM(z,\zeta)\varepsilon(\zeta,t)\d\zeta 
\end{equation}
with the kernel $M(z,\zeta) \in \mathbb{R}^{n \times n}$ is needed. This transformation removes the couplings in $H_0(z)\varepsilon(0,t)$ so that the resulting matrix premultiplying the local term becomes strictly lower triangular. Hence, a series connection of parabolic PDEs is obtained in the target system.
In particular, \eqref{trafc2} maps \eqref{drsc1} into the \emph{second intermediate target system}
 \begin{subequations}\label{drsc2}
 	\begin{align}
 	\partial_t\tilde{\varepsilon}(z,t) &= \Lambda(z)\partial_z^2\tilde{\varepsilon}(z,t) - \mu_c\tilde{\varepsilon}(z,t) - \tilde{A}_0(z)\tilde{\varepsilon}(0,t)\label{pdesc2}\\
 	\partial_z\tilde{\varepsilon}(0,t) &= 0\label{bc1c2}\\
 	\partial_z\tilde{\varepsilon}(1,t) &= \bar{Q}_1\tilde{\varepsilon}(1,t) + \textstyle\int_0^1K_2(\zeta)\tilde{\varepsilon}(\zeta,t)\d\zeta  + \tilde{C}_1w_0(t)\nonumber\\
 	& \quad   +  C_1w_1(t)\label{bc2c2}\\
 	\dot{w}_0(t) &= \tilde{F}_0w_0(t) + B_0\tilde{\varepsilon}(0,t)\label{plantodec3}\\
 	\dot{w}_1(t) &= F_1w_1(t) + \tilde{B}_0w_0(t) + Bu(t) \nonumber \\
 	& \quad + B_1\tilde{\varepsilon}(1,t) + \textstyle\int_0^1\bar{B}_1(\zeta)\tilde{\varepsilon}(\zeta,t)\d \zeta,\label{w1sys2}
 	\end{align}
 \end{subequations}
in which
\begin{subequations}
\begin{align}
 \bar{Q}_1    &= \tilde{Q}_1 - M(1,1)\\
 \bar{K}_2(\zeta) &=  K_1(\zeta) - M_z(1,\zeta)\label{firsteq}\\
 K_2(\zeta) &=  \bar{Q}_1M_I(1,\zeta) \! + \! \bar{K}_2(\zeta) \! + \! \int_{\zeta}^1\!\bar{K}_2(\bar{\zeta})M_I(\bar{\zeta},\zeta)\d\bar{\zeta}\\
 \bar{B}_1(\zeta) &= B_1\big(M_I(1,\zeta) \!+\! K_I(1,\zeta) \!+\!\! 	\int_{\zeta}^1\!\!K_I(1,\bar{\zeta})M_I(\bar{\zeta},\zeta)\d\bar{\zeta}\big).\label{lasteq}
\end{align}
\end{subequations}
Therein, $M_I(z,\zeta) \in \mathbb{R}^{n \times n}$ is the kernel determining the \emph{inverse backstepping transformation}
\begin{equation}\label{trafc2i}
 \varepsilon(z,t) = \mathcal{T}_{c,2}^{-1}[\tilde{\varepsilon}(t)](z) = \tilde{\varepsilon}(z,t) +  \int_0^zM_I(z,\zeta)\tilde{\varepsilon}(\zeta,t)\d\zeta. 
\end{equation}
Furthermore, the equations \eqref{firsteq}--\eqref{lasteq} result from similar calculations as for \eqref{K1def}--\eqref{lasteqimed}. Following the same derivation as for \eqref{keq}, the \emph{kernel equations} for \eqref{trafc2} are
\begin{subequations}\label{keq2}
	\begin{align}
	& \Lambda(z) M_{zz}(z,\zeta) - (M(z,\zeta)\Lambda(\zeta))_{\zeta\zeta}  
	= 0\label{kpde2}\\
	& M_{\zeta}(z,0)\Lambda(0) + M(z,0)\Lambda'(0) = \tilde{A}_0(z)\nonumber\\
	& \quad - H_0(z) + \textstyle\int_0^zM(z,\zeta)H_0(\zeta)\d\zeta\label{kBC12}\\
	& \Lambda(z) M'(z,z) + \Lambda(z)M_z(z,z) + M_{\zeta}(z,z)\Lambda(z) \nonumber\\
	&\quad  + M(z,z)\Lambda'(z)  = 0\label{kBC22}\\
	& M(z,z)\Lambda(z) - \Lambda(z)M(z,z) = 0\label{kBC32}\\  
	& M(0,0) = 0,\label{kIC2}
	\end{align}
\end{subequations}
in which \eqref{kpde2} is defined on $0 < \zeta < z < 1$. These kernel equations coincide with the ones found in \cite{Deu17}. Hence, the results in \cite{Deu17} ensure that \eqref{keq2} has a piecewise $C^2$-solution. This solution also determines the elements $\tilde{a}_{ij}(z)$, $i > j$,  in \eqref{a01def}.

As the PDE subsystem \eqref{pdesc2}--\eqref{bc2c2}  consists of a series connection of $n$ parabolic PDEs, a suitable choice of $\mu_c$ always exists to ensure its stability (see Section \ref{sec:clstab} for further details).

\begin{rem}
In principle, one can use the direct transformation $\tilde{\varepsilon}(z,t) = x(z,t) - \textstyle\int_0^zK(z,\zeta)x(\zeta,t)\d\zeta - N(z)w_0(t)$ to map \eqref{drs} into \eqref{drsc2}. Then, however, the kernel equations for $K(z,\zeta)$ and the IVP for $N(z)$ become mutually coupled. This significantly impedes their solution.	  \hfill $\triangleleft$
\end{rem}

\subsection{Decoupling Form for the $\Sigma_{n_1}$-System}\label{sec:decouplw1}
The crucial point of the backstepping design for DPS with an actuating ODE \eqref{plantode2} is to represent the latter into a particular form, which makes the decoupling into a stable ODE-PDE-ODE cascade possible. For this, introduce the change of coordinates
\begin{align}\label{ccdecoupl}
\bar{w}_1(t) = 
\begin{bmatrix}
\bar{w}_{1,1}(t)\\
\bar{w}_{1,2}(t) 
\end{bmatrix}
=
\underbrace{\begin{bmatrix}
 C_1\\
 T_1
\end{bmatrix}}_{T}w_1(t),
\end{align}
with $C_1 \in \mathbb{R}^{n \times n_1}$ and $T_1 \in \mathbb{R}^{n_1-n \times n_1}$ subject to the conditions
\begin{equation}\label{trafocond}
 \det T \neq 0 \quad \text{and} \quad T_1B = 0.
\end{equation}
The first condition in \eqref{trafocond} ensures that \eqref{ccdecoupl} qualifies as a change of coordinates. The second condition is imposed so that the input $u$ does not appear in the state equation for $\bar{w}_{1,2}(t)$. This is a prerequisite for the state feedback design. With $\operatorname{rk}C_1 = n$ and a vector relative degree of one for $(C_1,F_1,B)$  both implied by Assumption (A\ref{deouplu}), the results in \cite[Prop. 5.1.2]{Is95} ensure the existence of $T$ satisfying \eqref{trafocond} (see also \cite{Deu18a}). Then, a simple calculation utilizing \eqref{w1sys2}, leads to the \emph{decoupling form}
\begin{subequations}\label{epssys}
	\begin{align}
	\dot{\bar{w}}_{1,1}(t) &= F_{11}\bar{w}_{1,1}(t) + F_{12}\bar{w}_{1,2}(t) + C_1Bu(t)\label{eadynsfeed} \\
	                  & \qquad + P_{10}\tilde{\varepsilon}(1,t) + \textstyle\int_0^1P_{11}(\zeta)\tilde{\varepsilon}(\zeta,t)\d \zeta + P_{12}w_0(t)\nonumber\\
	\dot{\bar{w}}_{1,2}(t) &= F_{21}\bar{w}_{1,1}(t) + F_{22}\bar{w}_{1,2}(t)\label{biintdyn}\\
	& \qquad + P_{20}\tilde{\varepsilon}(1,t) + \textstyle\int_0^1P_{21}(\zeta)\tilde{\varepsilon}(\zeta,t)\d \zeta + P_{22}w_0(t), \nonumber
	\end{align}	
\end{subequations}
where 
\begin{equation}\label{Fmat}
\begin{bmatrix}
F_{11} & F_{12}\\
F_{21} & F_{22}
\end{bmatrix} = TF_1T^{-1} 
\end{equation}
and
\begin{subequations}
	\begin{alignat}{3}
	 &P_{10} = C_1B_1,  \quad   &&P_{11}(\zeta) = C_1\bar{B}_1(\zeta), \quad &&P_{12} = C_1\tilde{B}_0\\	 
	 &P_{20} = T_1B_1, \quad &&P_{21}(\zeta) = T_1\bar{B}_1(\zeta),  \quad &&P_{22} = T_1\tilde{B}_0.
	\end{alignat}	 
\end{subequations}

\subsection{Decoupling and Stabilization of the $\Sigma_{n_1}$-System}\label{sec:sffinal}
In the final decoupling step, the system \eqref{pdesc2}--\eqref{plantodec3}, \eqref{epssys} is mapped into the stable ODE-PDE-ODE cascade \eqref{drsc}. For this, introduce the new state 
\begin{equation}
 \tilde{w}_1(t) = 
 \begin{bmatrix}
  \tilde{w}_{1,1}(t)\\
  \tilde{w}_{1,2}(t)
 \end{bmatrix} = \mathcal{T}_{c,3}[w_0(t),\bar{w}_1(t),\tilde{\varepsilon}(t)]
\end{equation}
for \eqref{epssys} by the \emph{decoupling transformation}
\begin{subequations}\label{decouplstate}
	\begin{align}
	 \tilde{w}_{1,1}(t) &= \bar{w}_{1,1}(t)  \! - \! \Gamma_1w_0(t) \! - \!  \int_0^1\Sigma_1(\zeta)\tilde{\varepsilon}(\zeta,t)\d \zeta \! - \!  \Sigma_0\tilde{\varepsilon}(1,t) \label{omtraf1}\\
	 \tilde{w}_{1,2}(t) &= \bar{w}_{1,2}(t) - \Gamma_2w_0(t) - \int_0^1\Sigma_2(\zeta)\tilde{\varepsilon}(\zeta,t)\d\zeta\label{omtraf2}
	\end{align}
\end{subequations}
with $\Gamma_1 \in \mathbb{R}^{n \times n_0}$, $\Sigma_1(z) \in \mathbb{R}^{n \times n}$, $\Sigma_0 \in \mathbb{R}^{n \times n}$, $\Gamma_2 \in \mathbb{R}^{n_1 -n \times n_0}$ and $\Sigma_2(z) \in \mathbb{R}^{n_1 -n \times n}$. 

The transformation \eqref{omtraf1} is determined, in order to ensure the BC 
\begin{equation}
 \partial_z\tilde{\varepsilon}(1,t) = \tilde{w}_{1,1}(t)
\end{equation}
of the ODE-PDE-ODE cascade in \eqref{bc1c}. By taking  $\bar{w}_{1,1}(t) = C_1w_1(t)$ into account (see \eqref{ccdecoupl}), a comparison of the right-hand side of \eqref{bc2c2} with \eqref{omtraf1} directly yields
\begin{equation}\label{decouplpara}
 \Gamma_1 = -\tilde{C}_1, \quad \Sigma_1(\zeta) = -K_2(\zeta) \quad \text{and} \quad  \Sigma_0 = -\bar{Q}_1. 
\end{equation}
Consider the \emph{target system}
\begin{equation}
\dot{\tilde{w}}_{1,1}(t) = - K_{11}\tilde{w}_{1,1}(t) - K_{12}\tilde{w}_{1,2}(t)
\end{equation}
for \eqref{eadynsfeed}, in which $K_{11} \in \mathbb{R}^{n \times n}$ and $K_{12} \in \mathbb{R}^{n \times n_1 - n}$ are freely assignable \emph{feedback gains} (see \eqref{plantode2c}). Then, differentiation of \eqref{omtraf1} determined by \eqref{decouplpara} and making use of \eqref{drsc2} gives
\begin{align}
 u(t) &= (C_1B)^{-1}\Big((\Gamma_1\tilde{F}_0 - P_{12})w_0(t) - [F_{11} \;\; F_{12}]\bar{w}_1(t) 
        \nonumber\\
      &\quad 
      - [K_{11} \;\; K_{12}]\tilde{w}_{1}(t) \nonumber\\ 
      &\quad  +\big(\Gamma_1B_0 - \textstyle\int_0^1\Sigma_1(\zeta)\tilde{A}_0(\zeta)\d\zeta\big)\tilde{\varepsilon}(0,t)  -P_{10}\tilde{\varepsilon}(1,t)\nonumber\\
      &\quad  + \textstyle\int_0^1\Sigma_1(\zeta)\Lambda(\zeta)\partial^2_{\zeta}\tilde{\varepsilon}(\zeta,t)\d\zeta  + \Sigma_0\dot{\tilde{\varepsilon}}(1,t) \nonumber\\
      &\quad  - \textstyle\int_0^1(\mu_c\Sigma_1(\zeta) + P_{11}(\zeta))\tilde{\varepsilon}(\zeta,t)\d\zeta\Big)
\end{align}
after some intermediate calculations. By applying an integration by parts and \eqref{decouplstate}, it is readily shown that the resulting state feedback controller is of the form
\begin{equation}\label{foppre}
 u(t) = \mathcal{\tilde{K}}[w_0(t),\bar{w}_1(t),\tilde{\varepsilon}(0,t),\tilde{\varepsilon}(1,t),\dot{\tilde{\varepsilon}}(1,t),\tilde{\varepsilon}(t)].
\end{equation}
This result can be expressed in terms of the original coordinates $w_0$, $w_1$ and $x$ using \eqref{trafc2}, \eqref{decoupbacktrafo} and \eqref{ccdecoupl} to obtain the state feedback
\begin{equation}\label{sfeed}
 u(t) = \mathcal{K}[w_0(t),w_1(t),x(0,t),x(1,t),\dot{x}(1,t),x(t)]
\end{equation}
in the original coordinates.
\begin{rem}\label{remnoisediff}
The fact that the state feedback controller \eqref{foppre} depends on boundary values and a time derivative thereof is not unexpected. A similar result is obtained in \cite{Sab17,Deu18a,Wa18,Wa19} for the control of ODE-PDE-ODE systems. It should be remarked, however, that the implementation of this state feedback controller poses no serious problems in applications. In particular, systematic methods exist for the implementation of the derivative by making use of algebraic numeric differentiators (see, e.\:g., \cite{Kil13}). Since \eqref{sfeed} is implemented by an observer designed in Section \ref{sec:obs}, measurement noise is filtered by the observer and thus does not significantly influence the derivative estimation.
Furthermore, it should be noted that \eqref{foppre} is well-defined, which is ensured by the smooth solution of parabolic systems (see \cite {Deu18}). In particular, it is verified in the proof of Theorem \ref{thm:sloop} that the closed-loop system gives rise to an analytic $C_0$-semigroup so that the closed-loop solution is smooth both w.r.t. time and space. It is interesting to remark that the time derivative vanishes for Dirichlet actuation. Hence, it is needed to remove the local term at $z = 1$ in the corresponding Robin BC. \hfill $\triangleleft$
\end{rem}

Since \eqref{biintdyn} still depends on the states $\tilde{\varepsilon}(z,t)$ and $w_0(t)$, the second transformation \eqref{omtraf2} is introduced. Then, the latter states can be removed in the resulting $\tilde{w}_{1,1}$-system so that the decoupled system \eqref{plantode2c} is obtained. For this, assign the \emph{target system}
\begin{equation}\label{decoupltsys2}
\dot{\tilde{w}}_{1,2}(t) = \tilde{F}_{21}\tilde{w}_{1,1}(t) +  F_{22}\tilde{w}_{1,2}(t)
\end{equation}
for \eqref{biintdyn} (see \eqref{plantode2c}). In order to determine the corresponding transformation \eqref{omtraf2}, differentiate the latter, insert \eqref{pdesc2}--\eqref{bc2c2} and \eqref{epssys} in the result and use \eqref{decoupltsys2}. An integration by parts and some intermediate algebraic manipulations show that $\Sigma_2(z)$ in \eqref{omtraf2} has to be the solution of the BVP
\begin{subequations}\label{firstIVP}
	\begin{align}
  &(\Sigma_2\Lambda)''(z)  -  \mu_c\Sigma_2(z) - F_{22}\Sigma_2(z)\nonumber\\
  & \quad   =   P_{21}(z) - F_{21}K_2(z), \quad z \in (0,1)\label{odedecw1}\\
	&(\Sigma_2\Lambda)'(0) = \textstyle\int_0^1\Sigma_2(\zeta)\tilde{A}_0(\zeta)\d \zeta  
	  - \Gamma_2B_0\\
	&(\Sigma_2\Lambda)'(1) = F_{21}\bar{Q}_1 - P_{20}.
    \end{align}	
\end{subequations}    
With the solution of \eqref{firstIVP}, the matrix 
\begin{equation}\label{F21sol}
\tilde{F}_{21} = F_{21} - \Sigma_2(1)\Lambda(1)
\end{equation}
is determined. Furthermore, $\Gamma_2$ has to solve the Sylvester equation    
 \begin{equation}\label{syleq}
     F_{22}\Gamma_2 - \Gamma_2\tilde{F}_0 = F_{21}\tilde{C}_1 - P_{22}
\end{equation}
(see \eqref{F0tildeg}). The next lemma clarifies the solvability of \eqref{firstIVP} and \eqref{syleq}.
\begin{Lemma}\label{lem:bvpsf}
If
\begin{equation}\label{bvpcondsf}
\sigma(F_{22}) \cap \sigma(\tilde{F}_0) = \emptyset 
\end{equation}	
holds, then  the Sylvester equation \eqref{syleq} is uniquely solvable.	
Let $\sigma_c$ be the spectrum of the backstepping target system \eqref{pdesc}--\eqref{bc2c}. If
\begin{equation}\label{bvpcondsf2}
\sigma(F_{22}) \cap \sigma_c = \emptyset,
\end{equation}
then the BVP \eqref{firstIVP} has a unique piecewise $C^2$-solution $\Sigma_2$.
\end{Lemma}	

For the proof see Appendix \ref{app:lem:bvpsf}.

With \eqref{F21sol} known, the feedback gains $K_{11}$ and $K_{12}$ stabilizing the $\tilde{w}_1$-system \eqref{plantode2c} can be determined provided that the conditions of the next lemma are satisfied. 
\begin{Lemma}\label{lem:contr}
Assume that
\begin{equation}\label{eigasssfcond}
 v_i\t(\Sigma_2\Lambda)(1) \neq v_i\t F_{21}
\end{equation}  
holds for all linearly independent left eigenvectors $v_i$ of $F_{22}$ w.r.t. all eigenvalues $\nu_i$, $i = 1,2,\ldots,l_1$, $l_1 \leq n$. Then, there exist matrices $K_{11} \in \mathbb{R}^{n \times n}$ and $K_{12} \in \mathbb{R}^{n \times n_1 - n}$ ensuring a Hurwitz matrix $\tilde{F}_1$ in \eqref{F1til}.
\end{Lemma}

The proof of this lemma directly follows from applying the PHB eigenvector tests (see, e.\:g., \cite[Th. 6.2-5]{KL80}).

\begin{rem}
It should be remarked that the condition \eqref{bvpcondsf} can always be fulfilled by a suitable eigenvalue assignment for $\tilde{F}_0 = F_0 - B_0K_0$ in view of Assumption \ref{contrass}. Since the backstepping target system \eqref{pdesc}--\eqref{bc2c} is a cascade of parabolic PDEs with Neumann BCs, its real spectrum satisfies $\lambda \leq -\mu_c$, $\forall \lambda \in \sigma_c$. Hence, there always exists a $\mu_c$ such that condition \eqref{bvpcondsf2} is satisfied, which requires only the calculation of the eigenvalues of $F_{22}$. The condition \eqref{eigasssfcond} is obviously fulfilled for $F_{22}$ being a Hurwitz matrix. This, in particular, implies that the system $(C_1,F_1,B)$ is \emph{minimum phase}, since \eqref{epssys} is in \emph{Byrnes-Isidori normal form} (see, e.\:g., \cite[Ch. 4.3]{Is95}). 
\hfill $\triangleleft$
\end{rem}

\subsection{Stability of the State Feedback Loop}\label{sec:clstab}
It is obvious that the stability of the ODE and PDE subsystems in \eqref{drsc} implies the stability of the ODE-PDE-ODE cascade. The next theorem makes this more precise and shows closed-loop stability in the original coordinates.  
\begin{thm}[Stability of the state feedback loop]\label{thm:sloop}
Let $\sigma_c$ be the spectrum of the backstepping target system \eqref{pdesc}--\eqref{bc2c} and assume that the feedback gains $K_0$ in \eqref{F0tildeg} and $K_{11}$, $K_{12}$ in \eqref{F1til} are chosen such that
\begin{equation}
 \alpha_{c,0} = \max_{\lambda \in \sigma(\tilde{F}_{0})}\operatorname{Re}\lambda < 0 \;\; \text{and} \;\; \alpha_{c,1} = \max_{\lambda \in \sigma(\tilde{F}_{1})}\operatorname{Re}\lambda < 0
\end{equation}  
as well as 
\begin{equation}
 \sigma(\tilde{F}_0) \cap \sigma_c = \emptyset \quad \text{and} \quad \sigma(\tilde{F}_{1}) \cap \sigma_c = \emptyset. 
\end{equation}
If $\alpha_{\tilde{\varepsilon}} = - \mu_c < 0$, then the closed-loop system resulting from applying the state feedback controller \eqref{sfeed} to \eqref{drs} is well-posed in the state space $X = \mathbb{C}^{n_0} \oplus \mathbb{C}^{n_1} \oplus (L_2(0,1))^n$ with the usual weighted inner product. The latter induces the norm
\begin{equation}\label{normdef}
 \|\cdot\| = (\|\cdot\|^2_{\mathbb{C}^{n_0}} + \|\cdot\|^2_{\mathbb{C}^{n_1}} + \|\cdot\|_{L^n_2}^2)^{\frac{1}{2}},
\end{equation}
in which $\|h\|_{L^n_2} = (\int_0^1\|\Lambda^{-\frac{1}{2}}(\zeta)h(\zeta)\|^2_{\mathbb{C}^n}\d \zeta)^{1/2}$. The system is exponentially stable in the norm \eqref{normdef} with the decay rate
\begin{equation}
 \alpha_c = -\max(\alpha_{c,0},\alpha_{c,1},\alpha_{\tilde{\varepsilon}}) > 0.
\end{equation}
In particular, the closed-loop state $x_{c}(t) = \operatorname{col}(w_0(t),w_1(t),x(t))$ with $x(t) = \{x(z,t), z \in [0,1]\}$ satisfies
\begin{equation}\label{thexpstabsfeed}
\|x_c(t)\| \leq M_{c}\e^{-(\alpha_c + c)t}
\|x_c(0)\|,\quad  t \geq 0,
\end{equation}
for all $x_c(0) \in \mathbb{C}^{n_0} \oplus \mathbb{C}^{n_1} \oplus (H^2(0,1))^n \subset X$ compatible with  the BCs of the state feedback loop, an $M_c \geq 1$ and any $c < 0$ such that $\alpha_c + c > 0$. 
\end{thm}

For the proof see Appendix \ref{app:{thm:sloop}}.

\section{Observer design}\label{sec:obs}
Consider the \emph{observer}
\begin{subequations}\label{obs}
\begin{align}
\partial_t\hat{x}(z,t) &= \Lambda(z)\partial^2_z\hat{x}(z,t) + A(z)\hat{x}(z,t)  +  L(z)
\Delta(t)\label{xeqo}\\
\partial_z\hat{x}(0,t) &= Q_0\hat{x}(0,t) \! + \! C_0\hat{w}_0(t) \! + \! L^0_0\Delta(t) \! + \! L^1_0\dot{\Delta}(t)\label{uabco}\\
\partial_z\hat{x}(1,t) &= Q_1\hat{x}(1,t) + C_1\hat{w}_1(t) + L_1\Delta(t)\label{aktbco}\\
\dot{\hat{w}}_0(t) &= F_0\hat{w}_0(t) + B_0\hat{x}(0,t) + M_0\Delta(t)\label{plantodeo}\\
\dot{\hat{w}}_1(t)  &= F_1\hat{w}_1(t)\! + \! B_1\hat{x}(1,t) \! +\! Bu(t) \!+\!  M_1 \Delta(t)\label{plantode2o}
\end{align}
\end{subequations}
for \eqref{drs}, with 
\begin{equation}\label{inno}
 \Delta(t) = y(t) - C\hat{w}_0(t)
\end{equation}
as well as the \emph{observer gains} $L(z), L^0_{0}, L_0^1, L_1 \in \mathbb{R}^{n \times n}$, $M_0 \in \mathbb{R}^{n_0 \times n}$, $M_1 \in \mathbb{R}^{n_1 \times n}$ to be determined. The ICs of \eqref{obs} are $\hat{x}(z,0) = \hat{x}_0(z) \in \mathbb{R}^{n}$, $z \in [0,1]$, $\hat{w}_0(0) = \hat{w}_{0,0} \in \mathbb{R}^{n_0}$ and $\hat{w}_1(0) = \hat{w}_{1,0} \in \mathbb{R}^{n_1}$.
\begin{rem}
Similar to the observer design for hyperbolic ODE-PDE-ODE systems in \cite{Deu18a}, the proposed observer \eqref{obs} requires a time derivative of the output. This is necessary to obtain the corresponding backstepping target system for the observer error dynamics. This poses no problems in applications, because numerical differentiators can be utilized to implement these derivatives. More precisely, it is demonstrated in \cite{Kil13} that they can be parameterized to become non-sensitive to measurement noise. Note that the time derivative in the output injection is well-defined, since the observer error dynamics gives rise to an analytic $C_0$-semigroup (see proof of Theorem \ref{thm:obs}). Its effect on the resulting closed-loop performance needs further investigations and is a topic for future research. Similar to the state feedback design, the time derivative in \eqref{uabco} is not needed for a Dirichlet BC at $z = 0$. \phantom{leer}\hfill $\triangleleft$
\end{rem}

The design of the observer \eqref{obs} is based on the related \emph{observer error dynamics}. They results from defining  $e_x = x - \hat{x}$, $e_0 = w_0 - \hat{w}_0$, $e_1 = w_1 - \hat{w}_1$ and taking \eqref{drs}, \eqref{obs} into account. Then, by introducing the \emph{auxiliary observer gain} $\tilde{L}_0^1$ and setting
\begin{equation}\label{L01}
 L_0^1 = (Q_0 - \tilde{L}_0^1)(CB_0)^{-1}
\end{equation}
this yields the coupled linear parabolic ODE-PDE-ODE system%
\begin{subequations}\label{obse}
\begin{align}
 \partial_te_x(z,t) &= \Lambda(z)\partial^2_ze_x(z,t) + A(z)e_x(z,t) -L(z)Ce_0(t)\label{xeqoe}\\
 \partial_ze_{x}(0,t) &= \tilde{L}_0^1e_x(0,t) + \tilde{L}_0^0e_0(t)\label{uabcoe}\\
 \partial_ze_{x}(1,t) &= Q_1e_{x}(1,t) + C_1e_1(t) - L_1Ce_0(t)\label{aktbcoe}\\
 \dot{e}_0(t) &= (F_0 - M_0C)e_0(t) + B_0e_{x}(0,t)\label{plantodeoe}\\
 \dot{e}_1(t) &= F_1e_1(t) - M_1Ce_0(t) + B_1e_{x}(1,t)\label{plantode2oe}
\end{align}
\end{subequations}
with
\begin{equation}
 \tilde{L}_0^0 = C_0 - L_0^1CF_0 - (L^0_0 - L_0^1CM_0)C.\label{obsgain0}
\end{equation}
In this section, the sequence of transformations shown in Figure \ref{fig:tseqobs} is determined, which maps \eqref{obse} into the stable  \emph{ODE-PDE-ODE cascade}
\begin{subequations}\label{obsef}
\begin{align}
 \dot{\tilde{e}}_1(t) &= \tilde{\bar{F}}_1\tilde{e}_1(t)\label{plantode2oef}\\
 \partial_t\varepsilon_x(z,t) &= \Lambda(z)\partial^2_z\varepsilon_x(z,t) - \mu_o\varepsilon_x(z,t)\label{xeqoef}\\
 \partial_z\varepsilon_x(0,t) &= 0\\
 \partial_z\varepsilon_{x}(1,t) &= -\textstyle\int_0^1\tilde{\bar{A}}_0(\zeta)\varepsilon_x(\zeta,t)\d\zeta + C_1\tilde{e}_1(t)\label{uabcoef}\\
  \dot{\tilde{e}}_0(t) &= \tilde{\bar{F}}_0\tilde{e}_0(t) + \bar{D}\varepsilon_{x}(0,t)\label{plantodeoef} 
\end{align}
\end{subequations}
with
\begin{subequations}
\begin{align}
\tilde{\bar{F}}_1 &= F_1 - \tilde{M}_1C_1\label{ftilbardef}\\
\tilde{\bar{A}}_{0}(\zeta) &= \begin{bmatrix} 0    & \tilde{\bar{a}}_{12}(\zeta) & \ldots  & \tilde{\bar{a}}_{1n}(\zeta)\\
\vdots  & \ddots        & \ddots  & \vdots\\
\vdots  & \ddots        & \ddots  & \tilde{\bar{a}}_{n-1\, n}(\zeta)\\
0       & \ldots        &  \ldots & 0
\end{bmatrix}\label{a01defo}\\
 \tilde{\bar{F}}_0 &= \begin{bmatrix}
 \bar{F}_{11}  & -\tilde{M}_{12}\\
 \tilde{\bar{F}}_{21} & -\tilde{\bar{M}}_{22}
 \end{bmatrix}\label{F1tilo}\\
 \bar{D} &= \begin{bmatrix}
 0\\
 I_n
 \end{bmatrix}
\end{align}	
\end{subequations}
and the matrices in \eqref{F1tilo} being introduced in Section \ref{sec:decouplobs}. Herein, the matrix $\tilde{M}_1$ ensures that the $\tilde{e}_1$-system is stable. Similar to the state feedback case, the DPS \eqref{xeqoef}--\eqref{uabcoef} is a cascade of stable parabolic PDEs for suitable $\mu_o$, because of the strictly upper triangular matrix \eqref{a01defo}. Finally, the observer gains $\tilde{M}_{12}$ and $\tilde{\bar{M}}_{22}$ ensure that the $\tilde{e}_0$-system is stable, i.\:e., $\tilde{\bar{F}}_0$ in \eqref{F1tilo} is a Hurwitz matrix. This results in the stable ODE-PDE-ODE cascade \eqref{obsef} for the observer error dynamics.  A stability proof for \eqref{obsef} and the observer error dynamics in the original coordinates concludes the section.
\begin{figure}[t]
	\begin{equation*}
	\boxed{\Sigma_{n_1}(e_1)} \underset{z=1}{\Leftrightarrow} \boxed{\Sigma_{\infty}(e_x)} \underset{z=0}{\Leftrightarrow}\boxed{\Sigma_{n_0}(e_0)}
	\end{equation*}
	\vspace{-0.1cm}
	\begin{equation*}
	\hspace{0.8cm}\textbf{Sec. \ref{sec:decouplobs} \& \ref{sec:btrafinf}:} \quad \Downarrow \quad \tilde{e}_0 = \bar{T}e_0,\; e_x = \mathcal{T}^{-1}_{o,11}[\tilde{e}_x] 
	\end{equation*}
	\begin{equation*}
	\hspace{4.1cm} \tilde{e}_1 = \mathcal{T}_{o,12}[\tilde{e}_0,e_1,\tilde{e}_x]
	\end{equation*}
	\vspace{-0.2cm}
	\begin{equation*}
	\boxed{\Sigma^s_{n_1}(\tilde{e}_1)} \underset{z=1}{\Rightarrow} \boxed{\Sigma_{\infty}(\tilde{e}_x)} \underset{z=0}{\Leftrightarrow}\boxed{\Sigma_{n_0}(\tilde{e}_0)}
	\end{equation*}
	\vspace{-0.1cm}
	\begin{equation*}
	\hspace{0.43cm}\textbf{Sec. \ref{sec:obs3}:} \quad \Downarrow \quad \tilde{e}_x = \mathcal{T}^{-1}_{o,2}[\bar{e}_x] 
	\end{equation*}
	\vspace{-0.2cm}
	\begin{equation*}
	\boxed{\Sigma^s_{n_1}(\tilde{e}_1)} \underset{z=1}{\Rightarrow} \boxed{\Sigma^s_{\infty}(\bar{e}_x)} \underset{z=0}{\Leftrightarrow}\boxed{\Sigma_{n_0}(\tilde{e}_0)}
	\end{equation*}
	\vspace{-0.1cm}
	\begin{equation*}
	\hspace{0.935cm}\textbf{Sec. \ref{sec:obs4}:} \quad \Downarrow \quad\, \varepsilon_x = \mathcal{T}_{o,3}[\tilde{e}_0,\bar{e}_x] 
	\end{equation*}
	\vspace{-0.2cm}
	\begin{equation*}
	\boxed{\Sigma^s_{n_1}(\tilde{e}_1)} \underset{z=1}{\Rightarrow} \boxed{\Sigma^s_{\infty}(\varepsilon_x)} \underset{z=0}{\Rightarrow}\boxed{\Sigma^s_{n_0}(\tilde{e}_0)}
	\end{equation*}
	\caption{Sequence of transformations mapping the observer error dynamics into a stable ODE-PDE-ODE cascade. The $\Sigma^s$-systems are stabilized.}\label{fig:tseqobs}
\end{figure}%

\subsection{Decoupling Form for the $\Sigma_{n_0}$-System}\label{sec:decouplobs}
Similar to the $w_1$-system in the state feedback design (see Section \ref{sec:decouplw1}), the  $e_0$-system \eqref{plantodeoe} is mapped into a decoupling form. This is necessary for the decoupling of the $e_1$-system into the first intermediate target system in Section \ref{sec:btrafinf}. To this end, define the new state $\tilde{e}_0(t)$ by the inverse transformation
\begin{equation}\label{ccdecouplo}
e_0(t) = \underbrace{\begin{bmatrix}T_0 & B_0\end{bmatrix}}_{\bar{T}}
\underbrace{
\begin{bmatrix}
\tilde{e}_{0,1}(t)\\
\tilde{e}_{0,2}(t)
\end{bmatrix}}_{\tilde{e}_0(t)}
\end{equation}
with $T_0 \in \mathbb{R}^{n_0 \times n_0-n}$ and $B_0 \in \mathbb{R}^{n_0 \times n}$, in which 
\begin{equation}\label{trafocondo}
 \det \bar{T} \neq 0 \quad \text{and} \quad CT_0 = 0
\end{equation}
has to hold. Here, the second condition is required so that the output injection only depends on $\tilde{e}_{0,2}(t)$. By transposing the matrices appearing in \eqref{trafocondo} and taking $\operatorname{rk}B_0 = n$ as well as  Assumption (A\ref{rone}) into account, the results in \cite[Prop. 5.1.2]{Is95} imply the existence of $\bar{T}$ satisfying \eqref{trafocond}. By making use of  \eqref{plantodeoe}, \eqref{ccdecouplo} and \eqref{trafocondo}, the \emph{decoupling form}
\begin{subequations}\label{epssyso}
	\begin{align}
	\dot{\tilde{e}}_{0,1}(t) &= \bar{F}_{11}\tilde{e}_{0,1}(t) - \tilde{M}_{12}\tilde{e}_{0,2}(t)\\
	\dot{\tilde{e}}_{0,2}(t) &= \bar{F}_{21}\tilde{e}_{0,1}(t) - \tilde{M}_{22}\tilde{e}_{0,2}(t)  + e_{x}(0,t)
	\end{align}	
\end{subequations}
is obtained, where 
\begin{equation}\label{Fmato}
\bar{F}_0 =
\begin{bmatrix}
\bar{F}_{11} & \bar{F}_{12}\\
\bar{F}_{21} & \bar{F}_{22}
\end{bmatrix} = \bar{T}^{-1}F_0\bar{T} .
\end{equation}
The observer gains follow from
\begin{subequations}\label{Mtildef}
	\begin{align}
     \tilde{M}_{12} &= M_{12} - \bar{F}_{12}\\
 	 \tilde{M}_{22} &= M_{22} - \bar{F}_{22}   
	\end{align}
\end{subequations}
and
\begin{equation}\label{Mmato}
\begin{bmatrix}
M_{12}\\
M_{22}
\end{bmatrix} = \bar{T}^{-1}M_0CB_0.
\end{equation}

\subsection{Decoupling and Stabilization of the $\Sigma_{n_1}$-system}\label{sec:btrafinf}
Consider the \emph{backstepping transformation}
\begin{subequations}\label{decouplobstwotraf}
\begin{equation}\label{obstrafo}
e_x(z,t) = \mathcal{T}_{o,11}^{-1}[\tilde{e}_x(t)](z) =  \tilde{e}_x(z,t) - \int_0^zP_I(z,\zeta)\tilde{e}_x(\zeta,t)\d\zeta
\end{equation}
with the kernel $P_I(z,\zeta) \in \mathbb{R}^{n \times n}$ and the \emph{decoupling transformation}
\begin{align}\label{obsdecouplchange}
\tilde{e}_1(t) &= \mathcal{T}_{o,12}[\tilde{e}_0(t),e_1(t),\tilde{e}_x(t)]\\
               &= e_1(t)  \! - \! \Pi_1\tilde{e}_{0,1}(t) \! - \! \Pi_2\tilde{e}_{0,2}(t) \! - \! \int_0^1R(\zeta)\tilde{e}_x(\zeta,t)\d \zeta\nonumber
\end{align}
\end{subequations}
with $\Pi_1 \in \mathbb{R}^{n_1 \times n_0-n}$, $\Pi_2 \in \mathbb{R}^{n_1 \times n}$ and $R(z) \in \mathbb{R}^{n_1 \times n}$. 
Similar to the state feedback design, the backstepping transformation \eqref{obstrafo} removes the term $A(z)e_x(z,t)$ in the PDE \eqref{xeqoe}. In addition, the decoupling transformation \eqref{obsdecouplchange} decouples the $e_1$-system \eqref{plantode2oe} in the new coordinates from the other subsystems. Consequently, these transformations map the error system \eqref{uabcoe}--\eqref{plantodeoe} and \eqref{epssyso} into the \emph{first intermediate target system}
\begin{subequations}\label{obserrsys2}
	\begin{align}
	\dot{\tilde{e}}_1(t) &= \tilde{\bar{F}}_1\tilde{e}_1(t)\label{e1tilsys}\\
	\partial_t\tilde{e}_x(z,t) &= \Lambda(z)\partial^2_z\tilde{e}_x(z,t) - \mu_o\tilde{e}_x(z,t)  -F(z)\tilde{e}_{x}(0,t) \nonumber\\
	&\quad  -G(z)\tilde{e}_{0,1}(t) -\tilde{L}(z)\tilde{e}_{0,2}(t)\label{opdes1}\\
	\partial_z\tilde{e}_{x}(0,t) &= \bar{L}_0^1\tilde{e}_{x}(0,t) \! + \! \tilde{C}_0T_0\tilde{e}_{0,1}(t) \! + \! \tilde{L}_0^0B_0\tilde{e}_{0,2}(t) \label{dobsue2}\\	                       
	\partial_z\tilde{e}_{x}(1,t) &= -\textstyle\int_0^1\bar{H}_0(\zeta)\tilde{e}_x(\zeta,t)\d \zeta\nonumber\\
	& \quad  + C_1\Pi_1\tilde{e}_{0,1}(t) + \tilde{L}_1\tilde{e}_{0,2}(t) + C_1\tilde{e}_1(t)\label{obsrb2e2}\\	
	\dot{\tilde{e}}_{0,1}(t) &= \bar{F}_{11}\tilde{e}_{0,1}(t) - \tilde{M}_{12}\tilde{e}_{0,2}(t)\\
	\dot{\tilde{e}}_{0,2}(t) &= \bar{F}_{21}\tilde{e}_{0,1}(t) -  \tilde{M}_{22}\tilde{e}_{0,2}(t)  +  \tilde{e}_{x}(0,t),
	\end{align}
\end{subequations}
in which 
\begin{subequations}\label{auxgains}
\begin{equation}\label{Ctil0}
 \tilde{C}_0 = C_0 - L_0^1CF_0,
\end{equation}	
the \emph{auxiliary observer gains}
\begin{align}
\tilde{L}(z) &= \big(\mathcal{T}_{o,11}[L](z)C \! + \! P(z,0)\Lambda(0)\tilde{L}_0^0\big)B_0\label{obsgain}\\
 \bar{L}_0^1 &= \tilde{L}_0^1 + P_I(0,0)\label{Lb01}\\
\tilde{L}_1 &= C_1\Pi_2 - L_1CB_0\label{obsgain1}
\end{align} 
\end{subequations}
 and
\begin{subequations}\label{FGdef}
 \begin{align}
   F(z) &= -\mathcal{T}_{o,11}[P_{I,\zeta}(\cdot,0)](z)\Lambda(0)\nonumber\\
        &\quad\; -P(z,0)(\Lambda'(0) - \Lambda(0)\bar{L}_0^1)\\
   G(z) &= P(z,0)\Lambda(0)\tilde{C}_0T_0
 \end{align}
\end{subequations}
are utilized. In the latter expression, the inverse of \eqref{obstrafo} is the backstepping transformation
\begin{equation}
 \tilde{e}_x(z,t) = \mathcal{T}_{o,11}[e_x(t)](z) = e_x(z,t) + \int_0^zP(z,\zeta)e_x(\zeta,t)\d\zeta 
\end{equation}
with the kernel $P(z,\zeta) \in \mathbb{R}^{n \times n}$. The results \eqref{auxgains} and \eqref{FGdef} are based on applying the reciprocity relation  $\mathcal{T}_{o,11}[P_{I}(\cdot,0)](z) =  P(z,0)$ (see Remark \ref{rem:recrel}). Furthermore, 
\begin{equation}\label{Hbar0def}
 \bar{H}_o(\zeta) = \bar{A}_0(\zeta) - C_1R(\zeta)
\end{equation}
with
\begin{equation}\label{h01def}
\bar{A}_0(\zeta) = \begin{bmatrix} 0    & \bar{a}_{12}(\zeta) & \ldots  & \bar{a}_{1n}(\zeta)\\
\vdots  & \ddots        & \ddots  & \vdots\\
\vdots  & \ddots        & \ddots  & \bar{a}_{n-1\, n}(\zeta)\\
0       & \ldots        &  \ldots & 0
\end{bmatrix}
\end{equation}
holds in \eqref{obsrb2e2}. In view of Assumption (A\ref{obsass}), the matrix $\tilde{M}_1$ can always be chosen such that $\tilde{\bar{F}}_1 = F_1 - \tilde{M}_1C_1$ in \eqref{e1tilsys} is a Hurwitz matrix.

By differentiating \eqref{obstrafo} w.r.t. time and making use of \eqref{xeqoe}--\eqref{aktbcoe} and \eqref{opdes1}--\eqref{obsrb2e2}, an integration by parts and the Leibniz differentiation rule show that the kernel $P_I(z,\zeta)$ required to obtain the target system \eqref{obserrsys2} is the solution of the \emph{kernel equations}
\begin{subequations}\label{obsbvp}
	\begin{align}
	&\Lambda(z)P_{I,zz}(z,\zeta) \!-\! (P_I(z,\zeta)\Lambda(\zeta))_{\zeta\zeta} = -(\mu_oI \!+\! A(z))P_I(z,\zeta)\label{ocdbvp1}\\
	&P_{I,z}(1,\zeta) - Q_1P_{I}(1,\zeta) = -\bar{A}_0(\zeta)\label{obsbvpbc}\\
	&\Lambda(z)P_I'(z,z) + \Lambda(z)P_{I,z}(z,z) + P_{I,\zeta}(z,z)\Lambda(z)\nonumber\\
	&\;  + P_I(z,z)\Lambda'(z) = \mu_oI + A(z)\\
	&P_I(z,z)\Lambda(z) - \Lambda(z)P_I(z,z) = 0\label{ocdbvp2}\\
	&P_I(1,1) = -Q_1
	\end{align}
\end{subequations}
with \eqref{ocdbvp1} defined on $0 < \zeta < z < 1$. It is shown in \cite{Deu18} that these kernel equations can be traced back to \eqref{keq}. This implies that \eqref{obsbvp} has a piecewise $C^2$-solution. From this, the elements $\bar{a}_{ij}(z)$, $i < j$, of $\bar{A}_0(\zeta)$ in \eqref{h01def} are obtained. After differentiating \eqref{obsdecouplchange}, taking \eqref{obserrsys2} into account, an integration by parts as well as some algebraic manipulations yield the IVP%
\begin{subequations}\label{decoupleta2}
	\begin{align}
	&(R\Lambda)''(z) - \mu_oR(z) - \tilde{\bar{F}}_1R(z) \nonumber\\
	& \quad = \tilde{M}_1\bar{A}_0(z) - B_1P_I(1,z), \quad z \in [0,1)\label{decoulode2}\\
	&(R\Lambda)(1) = \tilde{M}_1\\
	&(R\Lambda)'(1) = -B_1\label{decouplrb12}
	\end{align}
\end{subequations}	
for $R(z)$ (see \eqref{ftilbardef}). From this, the matrix 
\begin{equation}
	\Pi_2 = \textstyle\int_0^1R(\zeta)F(\zeta)\d \zeta + (R\Lambda)(0)\bar{L}_0^1 - (R\Lambda)'(0)
\end{equation}
can be obtained so that $\Pi_1$ follows from the Sylvester equation
\begin{multline}
	\tilde{\bar{F}}_1\Pi_1 - \Pi_1\bar{F}_{11} = \Pi_2\bar{F}_{21}\\  
	 -\textstyle\int_0^1R(\zeta)G(\zeta)\d \zeta - R(0)\Lambda(0)\tilde{C}_0T_0 \label{sylsf2}	 
\end{multline}
(see \eqref{ftilbardef}). After solving \eqref{decoupleta2} and \eqref{sylsf2}, the observer gain	  
	\begin{multline}\label{M1gain}
	M_1 = \tilde{M}_1L_1 +\big(\tilde{\bar{F}}_1\Pi_2 + \Pi_2\tilde{M}_{22} + \Pi_1\tilde{M}_{12}\\
	 + R(0)\Lambda(0)\tilde{L}_0^0B_0 + \textstyle\int_0^1R(\zeta)\tilde{L}(\zeta)\d \zeta\big)(CB_0)^{-1}
	\end{multline}
can be determined.

The next lemma clarifies the solvability of \eqref{decoupleta2} and \eqref{sylsf2}, which can be ensured by a suitable choice of $\tilde{M}_1$ in $\tilde{\bar{F}}_{1} = F_1 - \tilde{M}_1C_1$ in view of Assumption \ref{obsass}.
\begin{Lemma}\label{lem:bvpsfo}
The IVP \eqref{decoupleta2} has a unique piecewise $C^2$-solution $R$. Furthermore, if 
\begin{equation}\label{lyapsol}
 \sigma(\tilde{\bar{F}}_{1}) \cap \sigma(\bar{F}_{11}) = \emptyset 
\end{equation}
holds, then the Sylvester equation \eqref{sylsf2} is uniquely solvable.
\end{Lemma}	
\begin{IEEEproof}
The proof follows from introducing $\tilde{R}(z) = R(z)\Lambda(z)$ and making use of the change of coordinates $\bar{z} = 1 -z$ so that $\tilde{R}(1-\bar{z}) = \bar{R}(\bar{z})$. This yields an IVP for $\bar{R}(\bar{z})$ of the form \eqref{ivp:sf}. Hence, with the same reasoning as in the proof of Lemma \ref{lem:sflemdecoupl}, the existence of a unique piecewise $C^2$-solution is ensured. The unique solvability of the Sylvester equation \eqref{sylsf2} is implied by \eqref{lyapsol} and \cite[Ch. 12.5, Th. 2]{La85}.
\end{IEEEproof}

\subsection{Stabilization of the $\Sigma_{\infty}$-System}\label{sec:obs3}
Since the matrix $\bar{H}_0(\zeta)$ in \eqref{obsrb2e2} is not strictly upper triangular (see \eqref{Hbar0def}), the PDE subsystem \eqref{opdes1}--\eqref{obsrb2e2} still consists of coupled parabolic PDEs. In order to obtain a series connection of PDEs as required by the ODE-PDE-ODE cascade \eqref{obsef} (see \eqref{a01defo}), the \emph{backstepping transformation}
\begin{equation}\label{btrafoobs}
 \tilde{e}_x(z,t) = \mathcal{T}^{-1}_{o,2}[\bar{e}_x(t)](z)
                  =  \bar{e}_x(z,t) - \int_0^zS_I(z,\zeta)\bar{e}_x(\zeta,t)\d\zeta
\end{equation}
with the kernel $S_I(z,\zeta) \in \mathbb{R}^{n \times n}$ is utilized. The corresponding \emph{second intermediate target system} reads
\begin{subequations}\label{obserrsys22}
	\begin{align}
	\dot{\tilde{e}}_1(t) &= \tilde{\bar{F}}_1\tilde{e}_1(t)\label{e1tilsys2}\\
	\partial_t\bar{e}_x(z,t) &= \Lambda(z)\partial^2_z\bar{e}_x(z,t) - \mu_o\bar{e}_x(z,t) - \bar{F}(z)\bar{e}_{x}(0,t) \nonumber\\
	&\quad  -\bar{G}(z)\tilde{e}_{0,1}(t) -\bar{L}(z)\tilde{e}_{0,2}(t)\label{opdes21}\\
	\partial_z\bar{e}_{x}(0,t) &= \tilde{C}_0T_0\tilde{e}_{0,1}(t)  +  \tilde{L}_0^0B_0\tilde{e}_{0,2}(t) \label{dobsue22}\\	                       
	\partial_z\bar{e}_{x}(1,t) &= -\textstyle\int_0^1\tilde{\bar{A}}_0(\zeta)\bar{e}_x(\zeta,t)\d \zeta\nonumber\\
	& \quad  + C_1\Pi_1\tilde{e}_{0,1}(t) + \tilde{L}_1\tilde{e}_{0,2}(t) + C_1\tilde{e}_1(t)\label{obsrb2e22}\\	
	\dot{\tilde{e}}_{0,1}(t) &= \bar{F}_{11}\tilde{e}_{0,1}(t) - \tilde{M}_{12}\tilde{e}_{0,2}(t)\label{etil0sys2a}\\
	\dot{\tilde{e}}_{0,2}(t) &= \bar{F}_{21}\tilde{e}_{0,1}(t)  -  \tilde{M}_{22}\tilde{e}_{0,2}(t)  + \bar{e}_{x}(0,t),\label{etil0sys2b}
	\end{align}
\end{subequations}
in which 
\begin{align}
\bar{L}_0^1 &= -S_I(0,0)\label{Lb10}\\
\bar{L}(z) &= \mathcal{T}_{o,2}[\tilde{L}](z) + S(z,0)\Lambda(0)\tilde{L}_0^0B_0\label{obsgain2}
\end{align} 
are \emph{auxiliary observer gains} and
\begin{subequations}\label{zgrobs}
	\begin{align}
	\bar{F}(z) &= \mathcal{T}_{o,2}[F](z) -\mathcal{T}_{o,2}[S_{I,\zeta}(\cdot,0)](z)\Lambda(0)-S(z,0)\Lambda'(0)\label{Fbardef}\\
	\bar{G}(z) &= \mathcal{T}_{o,2}[G](z) + S(z,0)\Lambda(0)\tilde{C}_0T_0.\label{Gbardef}
	\end{align}
\end{subequations}
The inverse of \eqref{btrafoobs} utilized in \eqref{obsgain2} and \eqref{zgrobs} is the backstepping transformation
\begin{equation}
 \bar{e}_x(z,t) = \mathcal{T}_{o,2}[\tilde{e}_x(t)](z) = \tilde{e}_x(z,t) + \int_0^zS(z,\zeta)\tilde{e}_x(\zeta,t)\d\zeta 
\end{equation}
with $S(z,\zeta) \in \mathbb{R}^{n \times n}$. By applying the  reciprocity  relation the result $\mathcal{T}_{o,2}[S_{I}(\cdot,0)](z) =  S(z,0)$ is obtained (see Remark \ref{rem:recrel}), which is used in \eqref{obsgain2} and \eqref{zgrobs}.

By making use of the same calculations as for \eqref{obsbvp}, the kernel equations 
\begin{subequations}\label{obsbvp2}
	\begin{align}
	&\Lambda(z)S_{I,zz}(z,\zeta) - (S_I(z,\zeta)\Lambda(\zeta))_{\zeta\zeta} = 0\label{ocdbvp12}\\
	&S_{I,z}(1,\zeta) = \bar{H}_0(\zeta) - \textstyle\int_{\zeta}^1\bar{H}_0(\bar{\zeta})S_I(\bar{\zeta},\zeta)\d\bar{\zeta} - \tilde{\bar{A}}_0(\zeta)\label{obsbvpbc2}\\
	&\Lambda(z)S_I'(z,z) + \Lambda(z)S_{I,z}(z,z) + S_{I,\zeta}(z,z)\Lambda(z)\nonumber\\
	&\;  + S_I(z,z)\Lambda'(z) = 0\\
	&S_I(z,z)\Lambda(z) - \Lambda(z)S_I(z,z) = 0\label{ocdbvp22}\\
	&S_I(1,1) = 0
	\end{align}
\end{subequations}
result for \eqref{btrafoobs} with \eqref{ocdbvp12} defined on $0 < \zeta < z < 1$. Their solution also determines the elements $\tilde{\bar{a}}_{ij}(z)$, $i < j$, of $\tilde{\bar{A}}_0(\zeta)$ in \eqref{a01defo}. Since \eqref{obsbvp2} has the same structure as \eqref{obsbvp} and the additional integral term in \eqref{obsbvpbc2} poses no problem (see the general type of kernel equations found in \cite{Deu17}), there exists a piecewise $C^2$-solution of \eqref{obsbvp2}. 

\subsection{Decoupling of the $\Sigma_{\infty}$-System and Stabilization of the $\Sigma_{n_0}$-System}\label{sec:obs4}
In the final transformation step, the PDE subsystem in \eqref{opdes21}--\eqref{obsrb2e22} is decoupled from the $\tilde{e}_{0}$-system \eqref{etil0sys2a}--\eqref{etil0sys2b}, in order to obtain the ODE-PDE-ODE cascade \eqref{obsef}. This requires to introduce the change of coordinates
\begin{align}
 \varepsilon_x(z,t) &= \mathcal{T}_{o,3}[\tilde{e}_0(t),\bar{e}_x(t)](z)\nonumber\\
 &= \bar{e}_x(z,t) - \Gamma_1(z)\tilde{e}_{0,1}(t) - \Gamma_2(z)\tilde{e}_{0,2}(t)\label{obsdecoupl} 
\end{align}
with $\Gamma_1(z) \in \mathbb{R}^{n \times n_0-n}$ and $\Gamma_2(z) \in \mathbb{R}^{n \times n}$. 

Differentiating \eqref{obsdecoupl} w.r.t. time, inserting \eqref{opdes21}--\eqref{obsrb2e22} and \eqref{etil0sys2a}--\eqref{etil0sys2b} lead to the BVP
\begin{subequations}\label{decoupleta2o}
	\begin{align}
	&\Lambda(z)\Gamma_1''(z) \! - \! \mu_o\Gamma_1(z) \! - \! \Gamma_1(z)\bar{F}_{11} =  \Gamma_2(z)\bar{F}_{21} \! + \! \bar{G}(z)\label{decoulode2o}\\
	&\Gamma'_1(0) =  \tilde{C}_0T_0\\
	&\Gamma'_1(1) = C_1\Pi_1 - \textstyle\int_0^1\tilde{\bar{A}}_0(\zeta)\Gamma_1(\zeta)\d \zeta	
	\end{align}
\end{subequations}
with \eqref{decoulode2o} defined on $z \in (0,1)$,
\begin{equation}
\Gamma_2(z) = -\bar{F}(z)
\end{equation}
(see \eqref{Fbardef}) and $\bar{G}(z)$ given by \eqref{Gbardef}. From this, the result
\begin{subequations}
\begin{align}
 \tilde{\bar{F}}_{21} &= \bar{F}_{21} + \Gamma_1(0)\label{Ftilb21}\\
 \tilde{\bar{M}}_{22} &= \tilde{M}_{22} - \Gamma_2(0)\label{Mtilb22}
\end{align}
\end{subequations}
is obtained. Finally, with \eqref{auxgains} and the solution of \eqref{decoupleta2o}, the observer gains are
\begin{subequations}\label{remaingains}
	\begin{align}
	L^0_0 &= \big(C_0B_0 - \Gamma'_2(0) - L_0^1CF_0B_0\big)(CB_0)^{-1} + L_0^1CM_0\label{L00}\\
	L_1 &= \big(C_1\Pi_2 - \Gamma'_2(1)  
	-\textstyle\int_0^1\tilde{\bar{A}}_0(\zeta)\Gamma_2(\zeta)\d\zeta\big)(CB_0)^{-1}\label{L1}\\
	L(z) &= \Big(\mathcal{T}_{o,11}^{-1}\mathcal{T}_{o,2}^{-1}[\Gamma_1\tilde{M}_{12} + \Gamma_2\tilde{M}_{22} + \Lambda\Gamma''_2 - \mu_o\Gamma_2](z)\nonumber\\
	& -\big(P_I(z,0) + \mathcal{T}_{o,11}^{-1}[S_I(\cdot,0)](z)\big)\Lambda(0)\tilde{L}_0^0B_0\Big)(CB_0)^{-1}.\label{L}
	\end{align}
\end{subequations}

The condition for the solvability of \eqref{decoupleta2o} is stated in the next lemma.
\begin{Lemma}\label{lem:beobdecoupl}
Let $\sigma_o$ be the spectrum of the backstepping target system \eqref{xeqoef}--\eqref{uabcoef}. If
\begin{equation}\label{obscondfin}
	\sigma(\bar{F}_{11}) \cap \sigma_o = \emptyset,
\end{equation}
then the BVP \eqref{decoupleta2o} has a unique piecewise $C^2$-solution $\Gamma_1$.
\end{Lemma}	
The proof follows the same lines as the proof of Lemma \ref{lem:bvpsf} and, thus, is omitted.

The next lemma presents the conditions for the stabilization of the $\tilde{e}_0$-system \eqref{plantodeoef}.
\begin{Lemma}\label{lem:obse0}
Assume that
\begin{equation}\label{obscond}
 \Gamma_1(0)\bar{v}_i \neq -\bar{F}_{21}\bar{v}_i
\end{equation}  
holds for all linearly independent right eigenvectors $\bar{v}_i$ of $\bar{F}_{11}$ w.r.t. the eigenvalues $\bar{\nu}_i$, $i = 1,2,\ldots,l_0$, $l_0 \leq n_0-n$. Then, there exist matrices $\tilde{M}_{12} \in \mathbb{R}^{n_0-n \times n}$ and $\tilde{\bar{M}}_{22} \in \mathbb{R}^{n \times n}$ so that $\tilde{\bar{F}}_0$ in \eqref{F1tilo} is a Hurwitz matrix.
\end{Lemma}

This lemma can readily be proved by utilizing the PHB eigenvector tests found in, e.\:g., \cite[Th. 6.2-5]{KL80}.
\begin{rem}
It should be noted that the condition \eqref{obscondfin} can always be fulfilled by computing the eigenvalues of $\bar{F}_{11}$ and a suitable choice of $\mu_o$ in \eqref{obserrsys22}. This is due to the fact that the backstepping target system \eqref{xeqoef}--\eqref{uabcoef} is a cascade of parabolic PDEs with Neumann BCs and thus its real spectrum satisfies $\lambda \leq -\mu_o$, $\forall \lambda \in \sigma_o$. Obviously, the condition \eqref{obscond} is fulfilled for a Hurwitz matrix $\bar{F}_{11}$. This can easily be checked for the plant \eqref{drs} by evaluating \eqref{Fmato}.
\hfill $\triangleleft$
\end{rem}

\subsection{Stability of the Observer Error Dynamics}\label{sec:obsstab}
Similar to the state feedback design, the stability of the ODE and PDE subsystems in \eqref{obsef} leads to a stable ODE-PDE-ODE cascade for the observer error dynamics. This is the result of the next theorem.  
\begin{thm}[Stability of the observer error dynamics]\label{thm:obs}
Let $\sigma_o$ be the spectrum of the backstepping target system \eqref{xeqoef}--\eqref{uabcoef} and assume that the observer gains $\tilde{M}_1$ in \eqref{ftilbardef} and $\tilde{M}_{12}$, $\tilde{\bar{M}}_{22}$ in \eqref{F1tilo} are chosen such that
\begin{equation}
\alpha_{o,0} = \max_{\lambda \in \sigma(\tilde{\bar{F}}_{0})}\operatorname{Re}\lambda < 0 \;\; \text{and} \;\; \alpha_{o,1} = \max_{\lambda \in \sigma(\tilde{\bar{F}}_{1})}\operatorname{Re}\lambda < 0
\end{equation}  
as well as 
\begin{equation}
\sigma(\tilde{\bar{F}}_0) \cap \sigma_o = \emptyset \quad \text{and} \quad \sigma(\tilde{\bar{F}}_{1}) \cap \sigma_o = \emptyset. 
\end{equation}
If $\alpha_{\varepsilon_x} = - \mu_o < 0$, then 
the error dynamics for the designed observer \eqref{obs} is well-posed in the state space $X = \mathbb{C}^{n_0} \oplus \mathbb{C}^{n_1} \oplus (L_2(0,1))^n$ with the usual weighted inner product. Furthermore, the observer error dynamics \eqref{obse} is exponentially stable in the norm $\|\cdot\|$ (see Theorem \ref{thm:sloop}) with the decay rate
\begin{equation}
\alpha_o = -\max(\alpha_{o,0},\alpha_{o,1},\alpha_{\varepsilon_x}) > 0.
\end{equation}
In particular, the observer error $e_{o}(t) = \operatorname{col}(e_0(t),e_1(t),e_x(t))$ with $e_x(t) = \{e_x(z,t), z \in [0,1]\}$ satisfies
\begin{equation}\label{thexpstabsfeedo}
\|e_o(t)\| \leq M_{o}\e^{-(\alpha_o + c)t}	
\|e_o(0)\|,\quad  t \geq 0,
\end{equation}
for all $e_o(0) \in \mathbb{C}^{n_0} \oplus \mathbb{C}^{n_1} \oplus (H^2(0,1))^n \subset X$ compatible with  the BCs of the observer error dynamics, an $M_o \geq 1$ and any $c < 0$ such that $\alpha_o + c > 0$. 
\end{thm}

For the proof see Appendix \ref{app:theo:obs}.

\section{Observer-based compensator}\label{sec:obscomp}
By utilizing the state estimates of the observer \eqref{obs} in the state feedback controller \eqref{sfeed}, i.\:e.,
\begin{equation}\label{sfeedest}
 u(t) = \mathcal{K}[\hat{w}_0(t),\hat{w}_1(t),\hat{x}(0,t),\hat{x}(1,t),
 \dot{\hat{x}}(1,t),\hat{x}(t)],
\end{equation}
an \emph{observer-based compensator} is obtained. The next theorem shows that the \emph{separation principle} is satisfied for the resulting closed-loop system implying closed-loop stability.
\begin{thm}[Stability of the closed-loop system]\label{thm:cloop}\hfill	
Let the state feedback controller \eqref{sfeed} and the observer \eqref{obs} be designed according to Theorems \ref{thm:sloop} and \ref{thm:obs}. Then, the closed-loop system resulting from applying the observer-based compensator to \eqref{drs} is  well-posed in the state space $X_{\mathit{cl}} = \mathbb{C}^{n_0} \oplus \mathbb{C}^{n_1} \oplus (L_2(0,1))^n \oplus \mathbb{C}^{n_0} \oplus \mathbb{C}^{n_1} \oplus (L_2(0,1))^n$ with the usual weighted inner product. Furthermore, the closed-loop system is exponentially stable in the norm
\begin{equation}
 \|\cdot\|_{\mathit{cl}} = (\|\cdot\|^2 + \|\cdot\|^2)^{\frac{1}{2}}
\end{equation}
(see \eqref{normdef}) with the decay rate
\begin{equation}
\alpha_{\mathit{cl}} = \min(\alpha_{c},\alpha_{o}) > 0.
\end{equation}
In particular, the closed-loop state $x_{\mathit{cl}}(t) = \operatorname{col}(w_0(t),w_1(t),\linebreak  x(t),\hat{w}_0(t),\hat{w}_1(t),\hat{x}(t))$
satisfies
\begin{equation}\label{thexpstabsclfeed}
\|x_{\mathit{cl}}(t)\|_{\mathit{cl}} \leq M_{\mathit{cl}}\e^{-(\alpha_{\mathit{cl}} + c)t}
\|x_{\mathit{cl}}(0)\|_{\mathit{cl}},\quad  t \geq 0,
\end{equation}
for all $x_{\mathit{cl}}(0) \in \mathbb{C}^{n_0} \oplus \mathbb{C}^{n_1} \oplus (H^2(0,1))^n \oplus \mathbb{C}^{n_0} \oplus \mathbb{C}^{n_1} \oplus (H^2(0,1))^n \subset X_{\mathit{cl}}$ compatible with  the BCs of the closed-loop system, an $M_{\mathit{cl}} \geq 1$ and any $c < 0$ such that $\alpha_{\mathit{cl}} + c > 0$. 
\end{thm}

For the proof see Appendix \ref{app:thm:cloop}.

\section{Example}\label{sec:ex}
The results of the paper are demonstrated by a numerical example. For this, consider a system of the form \eqref{drs} with the parameters
\begin{subequations}
\begin{align}
\Lambda(z) &= 
\begin{bmatrix}2+2\text{sin}(\pi z) & 0 \\0 & 1.1+\text{sin}(\pi z)\\\end{bmatrix}\\
A(z) &= \begin{bmatrix}0.5+z & 1 \\0.5 & 2+z\\\end{bmatrix}
\end{align}	 
\end{subequations}
for the PDE. The corresponding BCs are described by the matrices
\begin{subequations}
\begin{align}
 Q_0 &= \begin{bmatrix}2.25 & 0 \\0 & 1\\\end{bmatrix}, \quad
 C_0 = \begin{bmatrix}0 & 1 & 0 \\1 & 0 & 1\\\end{bmatrix}\\
 Q_1 &= \begin{bmatrix}2 & 0 \\0 & 3\\\end{bmatrix}, \quad
 C_1 = \begin{bmatrix}1.5 & 0 & 1 \\0 & 2 & 0.5\\\end{bmatrix}.
\end{align}
\end{subequations}
This PDE system is unstable with its largest eigenvalue at $16.28$. The sensing ODE coupled to the PDE at $z = 0$ is determined by
\begin{subequations}
\begin{align}
F_0 &= \begin{bmatrix}0.4 & 0 & 0 \\0 & 0.75 & 0\\0 & 0 & 0.2\\\end{bmatrix}, \quad 
B_0 = \begin{bmatrix}1.5 & 0\\0 & 1\\1 & 1\\\end{bmatrix}\\
C &= \begin{bmatrix}1 & 0 & 0 \\0 & 1 & 0\\\end{bmatrix}.
\end{align}	 
\end{subequations}
Finally, the matrices characterizing the actuating ODE at $z = 1$ are
\begin{equation}
F_1 = \begin{bmatrix}0.3 & 0 & 0 \\0 & 0.2 & 0\\0 & 0 & 0.1\\\end{bmatrix}, \quad
B_1 = \begin{bmatrix}1 & 0\\0 & 0.75\\1 & 1\\\end{bmatrix}, \quad B = \begin{bmatrix}2 & 0\\0 & 1\\2 & 3\\\end{bmatrix}.
\end{equation}
It is readily verified that the Assumptions (A\ref{contrass})--(A\ref{rone}) are fulfilled for this ODE-PDE-ODE system.

\textbf{State feedback design.} In the first step, the kernel equations \eqref{keq} are solved for $\mu_c = 2.3$ by making use of the \emph{method of successive approximations} (see \cite{Deu17} for details), which is also applied to determine the solution of all other kernel equations. With the resulting kernel, the IVP \eqref{ivp:sf} is solved, where in view of Assumption (A\ref{contrass}) a stabilizing state feedback gain $K_0$  exists. The latter is determined such that $\sigma(\tilde{F}_0) = \sigma(F_0 - B_0K_0) = \{-4.2, -4, -4.8\}$. The corresponding fundamental matrix is computed by discretizing the
$\zeta$-coordinate and solving the ODEs resulting from the corresponding IVP on the interval $z \in [\zeta,1]$ using \textsc{Matlab}'s solver \texttt{ode15s}. This approach is also utilized for all other IVPs and BVPs in the sequel. Then, the second kernel equations \eqref{keq2} are solved. With the choice
$T_1 = [{-1} \;\; {-3} \;\; 1]$ satisfying \eqref{trafocond}, the change of coordinates \eqref{ccdecoupl} is determined. The resulting matrix $F_{22}$ (see \eqref{Fmat}) fulfils the conditions \eqref{bvpcondsf} and \eqref{bvpcondsf2} ensuring the solvability of  the Sylvester equation \eqref{syleq} and the BVP \eqref{firstIVP} so that $\tilde{F}_{21}$ follows from \eqref{F21sol}. As the solution of \eqref{firstIVP} satisfies \eqref{eigasssfcond}, stabilizing feedback gains $K_{11}$ and $K_{12}$ exist. With this, the feedback gains $K_{11}$ and $K_{12}$ are determined to ensure $\sigma(\tilde{F}_1) = \{-5.3, -5, -6\}$ (see \eqref{F1til}). Finally, the state feedback controller is obtained by evaluating \eqref{sfeed}.
\begin{figure}[t]
	\centering
	\begin{tikzpicture}
	\matrix{
		\begin{axis}[%
width=.35\columnwidth,
height=.3\columnwidth,
at={(0em,2.em)},
scale only axis,
xmin=0,
xmax=1,
tick align=inside,
ymin=0,
ymax=6,
zmin=-300,
zmax=200,
view={53}{30},
axis background/.style={fill=white},
ztick = {-150,0,150},
title = {$x_1(z,t)$},
every axis title/.append style={yshift=-1.5ex},
ylabel = {$t$},
xlabel = {$z$},
every axis y label/.append style={yshift=2.75mm,xshift=-1mm},
every axis x label/.append style={yshift=2mm,xshift=4.5mm},
]

\addplot3[%
surf,
draw opacity=1,shader=faceted,restrict y to domain = 0:6.1,faceted color=black,colormap={mymap}{[1pt] rgb(0pt)=(0.2422,0.1504,0.6603); rgb(1pt)=(0.25039,0.164995,0.707614); rgb(2pt)=(0.257771,0.181781,0.751138); rgb(3pt)=(0.264729,0.197757,0.795214); rgb(4pt)=(0.270648,0.214676,0.836371); rgb(5pt)=(0.275114,0.234238,0.870986); rgb(6pt)=(0.2783,0.255871,0.899071); rgb(7pt)=(0.280333,0.278233,0.9221); rgb(8pt)=(0.281338,0.300595,0.941376); rgb(9pt)=(0.281014,0.322757,0.957886); rgb(10pt)=(0.279467,0.344671,0.971676); rgb(11pt)=(0.275971,0.366681,0.982905); rgb(12pt)=(0.269914,0.3892,0.9906); rgb(13pt)=(0.260243,0.412329,0.995157); rgb(14pt)=(0.244033,0.435833,0.998833); rgb(15pt)=(0.220643,0.460257,0.997286); rgb(16pt)=(0.196333,0.484719,0.989152); rgb(17pt)=(0.183405,0.507371,0.979795); rgb(18pt)=(0.178643,0.528857,0.968157); rgb(19pt)=(0.176438,0.549905,0.952019); rgb(20pt)=(0.168743,0.570262,0.935871); rgb(21pt)=(0.154,0.5902,0.9218); rgb(22pt)=(0.146029,0.609119,0.907857); rgb(23pt)=(0.138024,0.627629,0.89729); rgb(24pt)=(0.124814,0.645929,0.888343); rgb(25pt)=(0.111252,0.6635,0.876314); rgb(26pt)=(0.0952095,0.679829,0.859781); rgb(27pt)=(0.0688714,0.694771,0.839357); rgb(28pt)=(0.0296667,0.708167,0.816333); rgb(29pt)=(0.00357143,0.720267,0.7917); rgb(30pt)=(0.00665714,0.731214,0.766014); rgb(31pt)=(0.0433286,0.741095,0.73941); rgb(32pt)=(0.0963952,0.75,0.712038); rgb(33pt)=(0.140771,0.7584,0.684157); rgb(34pt)=(0.1717,0.766962,0.655443); rgb(35pt)=(0.193767,0.775767,0.6251); rgb(36pt)=(0.216086,0.7843,0.5923); rgb(37pt)=(0.246957,0.791795,0.556743); rgb(38pt)=(0.290614,0.79729,0.518829); rgb(39pt)=(0.340643,0.8008,0.478857); rgb(40pt)=(0.3909,0.802871,0.435448); rgb(41pt)=(0.445629,0.802419,0.390919); rgb(42pt)=(0.5044,0.7993,0.348); rgb(43pt)=(0.561562,0.794233,0.304481); rgb(44pt)=(0.617395,0.787619,0.261238); rgb(45pt)=(0.671986,0.779271,0.2227); rgb(46pt)=(0.7242,0.769843,0.191029); rgb(47pt)=(0.773833,0.759805,0.16461); rgb(48pt)=(0.820314,0.749814,0.153529); rgb(49pt)=(0.863433,0.7406,0.159633); rgb(50pt)=(0.903543,0.733029,0.177414); rgb(51pt)=(0.939257,0.728786,0.209957); rgb(52pt)=(0.972757,0.729771,0.239443); rgb(53pt)=(0.995648,0.743371,0.237148); rgb(54pt)=(0.996986,0.765857,0.219943); rgb(55pt)=(0.995205,0.789252,0.202762); rgb(56pt)=(0.9892,0.813567,0.188533); rgb(57pt)=(0.978629,0.838629,0.176557); rgb(58pt)=(0.967648,0.8639,0.16429); rgb(59pt)=(0.96101,0.889019,0.153676); rgb(60pt)=(0.959671,0.913457,0.142257); rgb(61pt)=(0.962795,0.937338,0.12651); rgb(62pt)=(0.969114,0.960629,0.106362); rgb(63pt)=(0.9769,0.9839,0.0805)},mesh/rows=15, z buffer = sort]
table[point meta=\thisrow{c}] {x1.txt};
\end{axis}
		&
		\begin{axis}[%
width=.35\columnwidth,
height=.3\columnwidth,
at={(0em,2.em)},
scale only axis,
xmin=0,
xmax=1,
tick align=inside,
ymin=0,
ymax=6,
zmin=-120,
zmax=120,
view={53}{30},
axis background/.style={fill=white},
ztick = {-75,0,75},
title = {$x_2(z,t)$},
every axis title/.append style={yshift=-1.5ex},
ylabel = {$t$},
xlabel = {$z$},
every axis y label/.append style={yshift=2.75mm,xshift=-1mm},
every axis x label/.append style={yshift=2mm,xshift=4.5mm},
]

\addplot3[%
surf,
draw opacity=1,shader=faceted,restrict y to domain = 0:6.1,faceted color=black,colormap={mymap}{[1pt] rgb(0pt)=(0.2422,0.1504,0.6603); rgb(1pt)=(0.25039,0.164995,0.707614); rgb(2pt)=(0.257771,0.181781,0.751138); rgb(3pt)=(0.264729,0.197757,0.795214); rgb(4pt)=(0.270648,0.214676,0.836371); rgb(5pt)=(0.275114,0.234238,0.870986); rgb(6pt)=(0.2783,0.255871,0.899071); rgb(7pt)=(0.280333,0.278233,0.9221); rgb(8pt)=(0.281338,0.300595,0.941376); rgb(9pt)=(0.281014,0.322757,0.957886); rgb(10pt)=(0.279467,0.344671,0.971676); rgb(11pt)=(0.275971,0.366681,0.982905); rgb(12pt)=(0.269914,0.3892,0.9906); rgb(13pt)=(0.260243,0.412329,0.995157); rgb(14pt)=(0.244033,0.435833,0.998833); rgb(15pt)=(0.220643,0.460257,0.997286); rgb(16pt)=(0.196333,0.484719,0.989152); rgb(17pt)=(0.183405,0.507371,0.979795); rgb(18pt)=(0.178643,0.528857,0.968157); rgb(19pt)=(0.176438,0.549905,0.952019); rgb(20pt)=(0.168743,0.570262,0.935871); rgb(21pt)=(0.154,0.5902,0.9218); rgb(22pt)=(0.146029,0.609119,0.907857); rgb(23pt)=(0.138024,0.627629,0.89729); rgb(24pt)=(0.124814,0.645929,0.888343); rgb(25pt)=(0.111252,0.6635,0.876314); rgb(26pt)=(0.0952095,0.679829,0.859781); rgb(27pt)=(0.0688714,0.694771,0.839357); rgb(28pt)=(0.0296667,0.708167,0.816333); rgb(29pt)=(0.00357143,0.720267,0.7917); rgb(30pt)=(0.00665714,0.731214,0.766014); rgb(31pt)=(0.0433286,0.741095,0.73941); rgb(32pt)=(0.0963952,0.75,0.712038); rgb(33pt)=(0.140771,0.7584,0.684157); rgb(34pt)=(0.1717,0.766962,0.655443); rgb(35pt)=(0.193767,0.775767,0.6251); rgb(36pt)=(0.216086,0.7843,0.5923); rgb(37pt)=(0.246957,0.791795,0.556743); rgb(38pt)=(0.290614,0.79729,0.518829); rgb(39pt)=(0.340643,0.8008,0.478857); rgb(40pt)=(0.3909,0.802871,0.435448); rgb(41pt)=(0.445629,0.802419,0.390919); rgb(42pt)=(0.5044,0.7993,0.348); rgb(43pt)=(0.561562,0.794233,0.304481); rgb(44pt)=(0.617395,0.787619,0.261238); rgb(45pt)=(0.671986,0.779271,0.2227); rgb(46pt)=(0.7242,0.769843,0.191029); rgb(47pt)=(0.773833,0.759805,0.16461); rgb(48pt)=(0.820314,0.749814,0.153529); rgb(49pt)=(0.863433,0.7406,0.159633); rgb(50pt)=(0.903543,0.733029,0.177414); rgb(51pt)=(0.939257,0.728786,0.209957); rgb(52pt)=(0.972757,0.729771,0.239443); rgb(53pt)=(0.995648,0.743371,0.237148); rgb(54pt)=(0.996986,0.765857,0.219943); rgb(55pt)=(0.995205,0.789252,0.202762); rgb(56pt)=(0.9892,0.813567,0.188533); rgb(57pt)=(0.978629,0.838629,0.176557); rgb(58pt)=(0.967648,0.8639,0.16429); rgb(59pt)=(0.96101,0.889019,0.153676); rgb(60pt)=(0.959671,0.913457,0.142257); rgb(61pt)=(0.962795,0.937338,0.12651); rgb(62pt)=(0.969114,0.960629,0.106362); rgb(63pt)=(0.9769,0.9839,0.0805)},mesh/rows=15, z buffer = sort]
table[point meta=\thisrow{c}] {x2.txt};
\end{axis}%
		\\
		\begin{axis}[%
width=0.35\columnwidth,
height=0.3\columnwidth,
at={(0em,2.em)},
scale only axis,
xmin=0,
xmax=6,
xmajorgrids,
ymin=-22,
ymax=22,
ymajorgrids,
axis background/.style={fill=white},
ytick={-15,0,15},
xlabel = {$t$},
title = {$w_0(t)$},
every axis title/.append style={yshift=-1.5ex},
every axis x label/.append style={yshift=1ex},
legend columns=1,
legend entries ={$w_{0,1}$,$w_{0,2}$,$w_{0,3}$},
legend to name = w0legend,
]
\addplot [color = black!40!blue,dashdotted,line width = 1 pt]
table[]{w01.txt};
\addplot [color = black!50!green,dashed,line width = 1 pt]
table[]{w02.txt};
\addplot [color = black!40!red,line width = 1 pt]
table[]{w03.txt};
\end{axis}%
		&
		\begin{axis}[%
width=0.35\columnwidth,
height=0.3\columnwidth,
at={(0em,2.em)},
scale only axis,
xmin=0,
xmax=6,
xmajorgrids,
ymin=-300,
ymax=320,
ymajorgrids,
axis background/.style={fill=white},
ytick={-200,0,200},
yticklabels={-2,0,2},
xlabel = {$t$},
title = {$w_1(t)$},
every axis title/.append style={yshift=-1.5ex},
every axis x label/.append style={yshift=1ex},
ylabel near ticks,
clip mode=individual,
]
\addplot [color = black!40!blue,dashdotted,line width = 1 pt]
table[]{w11.txt};\label{pgf:w11}
\addplot [color = black!50!green,dashed,line width = 1 pt]
table[]{w12.txt};\label{pgf:w12}
\addplot [color = black!40!red,line width = 1 pt]
table[]{w13.txt};\label{pgf:w13}
\node[anchor = east] at (axis cs: 0.2,310) {$10^2$};
\end{axis}%
		\\
	};
	\end{tikzpicture}
	\vspace{-1em}
	\caption{Profiles of the distributed closed-loop states (upper plots) and lumped closed-loop states $w_{i,1}$ \eqref{pgf:w11}, $w_{i,2}$ \eqref{pgf:w12} and $w_{i,3}$ \eqref{pgf:w13} with $i=0,1$ (lower plots).}\label{fig:1}
\end{figure}
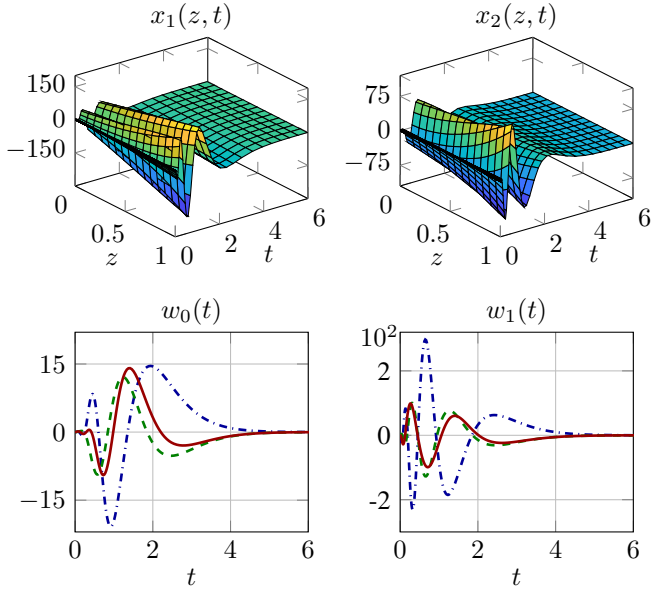

\textbf{Observer design.} In order to provide a systematic design procedure, the various steps to determine the observer gains $L(z)$, $L^0_{0}$, $L_0^1$ and $L_1$ for the PDE subsystem as well as $M_0$ and $M_1$ for the the ODE subsystems in \eqref{obs} are outlined:
\begin{itemize}
	\item[\circled{\small 1}] Determine $\tilde{M}_1$ such that $\sigma(\tilde{\bar{F}}_1) = \sigma(F_1 - \tilde{M}_1C_1) = \{-10.6,-10, -12\}$, which is possible due to Assumption (A\ref{obsass}).\\[-0.3cm]
	\item[\circled{\small 2}] Choose $\mu_o = 4.6$ and solve \eqref{obsbvp} as well as \eqref{decoupleta2}.\\[-0.3cm]	
	\item[\circled{\small 3}] Solve \eqref{obsbvp2} so that $S_I(z,\zeta) \xrightarrow{\eqref{Lb10}} \bar{L}_0^1 \xrightarrow{\eqref{Lb01}} \tilde{L}_0^1 \xrightarrow{\eqref{L01}} \bm{L_0^1} \xrightarrow{\eqref{Ctil0}} \tilde{C}_0$ can be computed.\\[-0.3cm]
	\item[\circled{\small 4}] Choose $T_0 = [0 \; 0 \; 1]\t$ satisfying \eqref{trafocondo} to get $\bar{F}_0$ from \eqref{Fmato}. Then, solve \eqref{decoupleta2o} and check \eqref{obscond}, which is fulfilled for the example. Hence, one can obtain $\tilde{\bar{F}}_{21}$ from \eqref{Ftilb21} so that $\tilde{M}_{12}$ and $\tilde{\bar{M}}_{22}$ are computed to achieve $\sigma(\tilde{\bar{F}}_0) = \{-8.2, -8,-9.6\}$ (see \eqref{F1tilo}). This gives $\tilde{\bar{M}}_{22} \xrightarrow{\eqref{Mtilb22}} \tilde{M}_{22}$, $\tilde{M}_{12}, \tilde{M}_{22} \xrightarrow{\eqref{Mtildef}} M_{12}, M_{22}  \xrightarrow{\eqref{Mmato}} \bm{M_0}$, and $\bm{L_1}$ results from \eqref{L1}.\\[-0.3cm]
	\item[\circled{\small 5}] Next, compute $L_0^1, M_0 \xrightarrow{\eqref{L00}} \bm{L_0^0}$ and consider $L_0^0, L_0^1, M_0 \xrightarrow{\eqref{obsgain0}} \tilde{L}_0^0$ as well as $\tilde{L}_0^0, \tilde{M}_{12}, \tilde{M}_{22} \xrightarrow{\eqref{L}} \bm{L(z)}$.\\[-0.3cm]
	\item[\circled{\small 6}] Finally, solve \eqref{sylsf2} and evaluate $\tilde{M}_1, L_1, \tilde{M}_{12}, \tilde{M}_{22},\linebreak \tilde{L}_0^0, \tilde{L}(z) \xrightarrow{\eqref{M1gain}} \bm{M_1}$ completing the observer design. \hfill $\Box$		
\end{itemize}

\textbf{Simulation.} The closed-loop system is simulated by making use of a FEM model with 151 grid points. The ICs of the plant are $x(z,0) = \operatorname{col}(5\left(\frac{3}{4}\text{sin}(\pi z+2\pi) +\frac{1}{4}\text{cos}(3\pi z+\frac{\pi}{2})\right),\linebreak 5\left(\frac{3}{4}\text{sin}(\pi z+2\pi) +\frac{1}{4}\text{cos}(3\pi z+\frac{\pi}{2})\right))$, $w_0(0) = 0$ and $w_1(0) = 0$, which are consistent to the BCs. Furthermore, all ICs of the compensator are set to zero. In Figure \ref{fig:1}, the resulting closed-loop response is shown  for the distributed states of the PDE subsystem (upper plots) and for the lumped states of the ODE subsystems (lower plot). This result confirms that the compensator ensures the stabilization of the closed-loop system for the considered ICs. In Figure \ref{fig:2}, the specification of the stability margin for the closed-loop system is verified. Obviously, exponential convergence w.r.t. the closed-loop norm is achieved and a larger stability margin can be assigned by increasing the design parameter $\mu_c$. The control inputs for $\mu_c = 4$ are depicted in Figure \ref{fig:3}. In comparison, the control inputs range between $-1$ and $1$ for $\mu_c = 2.3$ with a similar time response, which shows the increase of the control effort for a faster closed-loop system.

%
%
%
\begin{figure}[t]
	\centering
	\begin{tikzpicture}
	\matrix{
%
%
%
\begin{axis}[%
width=0.35\columnwidth,
height=0.3\columnwidth,
at={(0em,2.em)},
scale only axis,
xmin=0,
xmax=6,
xmajorgrids,
ymin=0,
ymax=550,
ymajorgrids,
axis background/.style={fill=white},
legend style={legend cell align=left,align=left,draw=white!15!black},
xlabel = {$t$},
ylabel = {$\|x_{cl}(t)\|_{cl}$},
title = {$\mu_c = 2.3$},
every axis title/.append style={yshift=-1.5ex},
every axis x label/.append style={yshift=1ex},
ytick={200,400},
yticklabels={2,4},
ylabel near ticks,
clip mode=individual,
]
\addplot [color=black!40!blue,solid,forget plot, line width = 1 pt]
table[]{Norm_mu2_3-240000_01.txt};
\addplot [color=black!40!red,dashed,forget plot,line width = 1 pt]
table[]{Norm_mu2_3-240000_02.txt};
\node[anchor = west] at (axis cs: 3,290)
{$c_1\text{e}^{\alpha_{\mathit{cl}}t}$}; 
\node[anchor = east] at (axis cs: 0.2,550) {$10^2$};
\end{axis}%
		&
%
%
%
\begin{axis}[%
width=0.35\columnwidth,
height=0.3\columnwidth,
at={(0em,2.em)},
scale only axis,
xmin=0,
xmax=6,
xmajorgrids,
ymin=0,
ymax=1400,
ymajorgrids,
axis background/.style={fill=white},
legend style={legend cell align=left,align=left,draw=white!15!black},
xlabel = {$t$},
title = {$\mu_c = 4$},
every axis title/.append style={yshift=-1.5ex},
every axis x label/.append style={yshift=1ex},
ytick = {500,1000},
yticklabels={5,10},
ylabel near ticks,
clip mode=individual,
]
\addplot [color=black!40!blue,solid, line width = 1 pt]
table[]{Norm_mu4_6000000_01.txt};\label{pgf:norm}
\addplot [color=black!40!red,dashed,forget plot,line width = 1 pt]
table[]{Norm_mu4_6000000_02.txt};\label{pgf:exp}
\node[anchor = west] at (axis cs: 2.4,725) {$c_2\text{e}^{\alpha_{\mathit{cl}}t}$};
\node[anchor = east] at (axis cs: 0.2,1400) {$10^2$};
\end{axis}%
		\\
	};
	\end{tikzpicture}
	\vspace{-1em}
	\caption{Norms of the closed-loop state $x_{\mathit{cl}}$ \eqref{pgf:norm} for two different $\mu_c$ and the corresponding exponential bounds \eqref{pgf:exp} with $c_1 = 2.4\cdot10^{5}$ and $c_2=6\cdot10^{6}$.}\label{fig:2}
\end{figure}
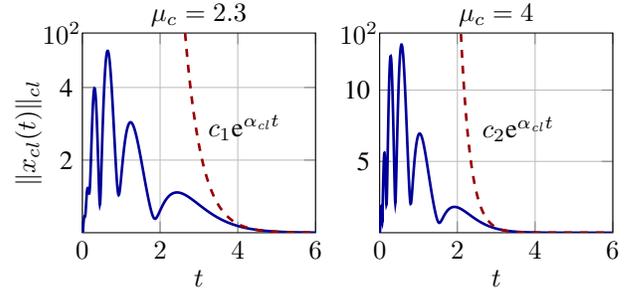
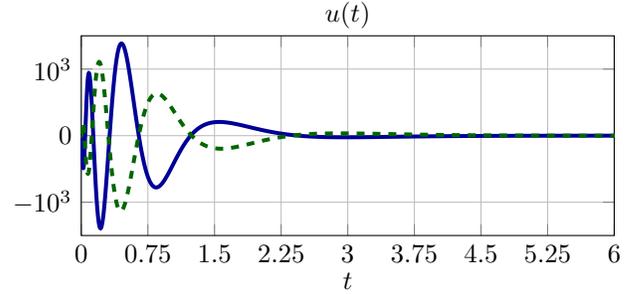
\begin{figure}
	\centering
	\begin{tikzpicture}
	\begin{axis}[%
	name = Uaxis,
	width=0.8\columnwidth,
	height=.3\columnwidth,
	anchor = outer north west,
	scale only axis,
	xmin=0,
	xmax=6,
	xmajorgrids,
	ymin=-1500,
	ymax=1500,
	ymajorgrids,
	axis background/.style={fill=white},
	title = {$u(t)$},
	xlabel = {$t$},
	xtick = {0,0.75,1.5,2.25,3,3.75,4.5,5.25,6},
	ytick={-1000,0,1000},
	yticklabels={$-10^3$,$0$,$10^3$},
	every axis title/.append style={yshift=-1.5ex},
	every axis x label/.append style={yshift=1ex},
	legend style={legend cell align=left,align=left,draw=white!15!black}
	]
	\addplot [color = black!40!blue,line width = 1.5 pt]
	table[]{u1.txt};\label{pgf:u1}
	\addplot [color = black!60!green,dashed,line width = 1.5 pt]
	table[]{u2.txt};\label{pgf:u2}
	\end{axis}%
	\end{tikzpicture}
	\caption{Plots of the controls $u_1(t)$ \eqref{pgf:u1} and $u_2(t)$ \eqref{pgf:u2} for  $\mu_c = 4$.}\label{fig:3}
\end{figure}

\section{Concluding remarks}
The proposed observer design can also be considered for a collocated setup, in which the actuating and sensing ODE coincide. By making use of the approach in \cite{Vaz16a}, an extension of the results to coupled diffusion-advection-reaction systems with an advection term of no particular form seems possible, which is of interest for continuing investigations. Real world applications may require to take nonlinear actuator and sensor dynamics into account. Hence, the generalization of the presented results to this class of nonlinear parabolic ODE-PDE-ODE systems is an interesting topic for further research.

%

\appendix

\subsection{Proof of Lemma \ref{lem:bvpsf}}\label{app:lem:bvpsf}
In order to represent \eqref{firstIVP} in a simple form, define $\tilde{\Sigma}_2(z) = \Sigma_2(z)\Lambda(z)$ to obtain
\begin{subequations}\label{bvp2}
	\begin{align}
	\tilde{\Sigma}_2''(z) &= (\mu_cI + F_{22})\tilde{\Sigma}_2(z)\Lambda^{-1}(z) + H(z)\label{dode12}\\
	\tilde{\Sigma}_2'(0)  &=  \textstyle\int_0^1\!\tilde{\Sigma}_2(\zeta)\Lambda^{-1}(\zeta)\tilde{A}_0(\zeta)\d \zeta  -  \Gamma_2B_0\label{dode22}\\
	\tilde{\Sigma}_2'(1) &= F_{21}\bar{Q}_1 - P_{20}\label{dode32}
	\end{align}
\end{subequations}
with $H(z) = P_{21}(z) - F_{21}K_2(z)$ and \eqref{dode12} defined on $z \in (0,1)$. Assume that the matrix $F_{22} \in \mathbb{R}^{n_1 - n \times n_1 -n}$ has $r$ \emph{Jordan blocks} $J_i$, $i = 1,2,\ldots,r$, to which the \emph{Jordan chains}%
\begin{subequations}\label{jchain}
	\begin{align}
	\varphi\t_{i(1)}F_{22} &= \mu_i\varphi\t_{i(1)}\label{Fetaewp}\\
	\varphi\t_{i(k)}F_{22} &= \mu_i\varphi\t_{i(k)} + \varphi\t_{i(k-1)}, \quad 
	k = 2,3,\ldots,l_i,
	\end{align}
\end{subequations}
are associated and $l_1 + \ldots + l_r = n_1 - n$. In \eqref{jchain} the vectors $\varphi_{i(k)}$ are the \emph{generalized left eigenvectors} of $F_{22}$ w.r.t. the eigenvalue $\mu_i$, $i = 1,2,\ldots,r$, in which $r$ is the number of eigenvalues of $F_{22}$  with linearly independent eigenvectors  (see, e.\:g., \cite[Ch. 6]{La85} for details on the related \emph{Jordan canonical form}). In the following it is assumed that the generalized eigenvectors $\varphi_{i(k)}$, $i = 1,2,\ldots,r$, $k = 1,2,\ldots,l_i$, are determined such that they form a basis of $\mathbb{C}^{n_1-n}$, which is always possible. Premultiplying \eqref{bvp2} by the generalized eigenvectors $\varphi\t_{i(k)}$, defining $\sigma_{i(k)}\t(z) = \varphi_{i(k)}\t \tilde{\Sigma}_2(z)$, $h_{i(k-1)}\t(z) = \sigma\t_{i(k-1)}(z) + \varphi_{i(k)}\t H(z)$, $\gamma_{i(k)}\t = \varphi_{i(k)}\t\Gamma_2B_0$ and $p_{i(k)}\t = \varphi_{i(k)}\t(F_{21}\bar{Q}_1 - P_{20})$ lead to the one-sided coupled BVPs
\begin{subequations}\label{bvp2m}
	\begin{align}
	\d_z^2\sigma\t_{i(k)}(z) &= (\mu_c + \mu_i)\sigma\t_{i(k)}(z)\Lambda^{-1}(z) + h_{i(k-1)}\t(z)\label{dode12m}\\
	\d_z\sigma\t_{i(k)}(0) &=  \textstyle\int_0^1\sigma\t_{i(k)}(\zeta)\Lambda^{-1}(\zeta)\tilde{A}_0(\zeta)\d \zeta  -  \gamma_{i(k)}\t\label{dode22m}\\
	\d_z\sigma\t_{i(k)}(1) &= p_{i(k)}\t\label{dode32m}
	\end{align}
\end{subequations}
for $i = 1,2,\ldots,r$, $k = 1,2,\ldots,l_i$, and $\sigma_{i(0)}\t(z) \equiv 0$. In what follows, the ODE \eqref{dode12m} is reformulated as a state space representation. The particular choice  $\rho_{ik,1}\t = -\d_z\sigma_{i(k)}\t$ and $\rho_{ik,2}\t = \sigma_{i(k)}\t$ of the state variables ensures that the resulting IVP will be adjoint to the IVP resulting from the eigenvalue problem related to the backstepping target system \eqref{pdesc}--\eqref{bc2c} (see below). After introducing the matrices 
\begin{equation}\label{Ematdef}
 E_1 = \begin{bmatrix}
I_{n_1-n}\\
0
\end{bmatrix} \quad \text{and} \quad
E_2 = \begin{bmatrix}
0\\
I_{n_1-n}
\end{bmatrix}
\end{equation}
with $E_1, E_2 \in \mathbb{R}^{2(n_1 -n) \times (n_1 -n)}$, the corresponding state space representation reads
\begin{equation}\label{statebvp}
\d_z\rho\t (z) = -\rho\t(z)\Upsilon(z,\mu_i) - h_{i(k-1)}\t(z)E_1\t
\end{equation}
with the state $\rho\t = [\rho_{ik,1}\t \;\; \rho_{ik,2}\t]$ and
\begin{equation}\label{Adef}
\Upsilon(z,s) = \begin{bmatrix}
0 & I\\
(s + \mu_c)\Lambda^{-1}(z) & 0
\end{bmatrix}.
\end{equation}
The \emph{fundamental matrix} $\Psi(z,\zeta,s) : (0,1)^2 \times \mathbb{C} \to \mathbb{C}^{2(n_1 - n) \times 2(n_1 -n)}$ related to \eqref{Adef} satisfies the IVP
\begin{equation}\label{Psidef}
\partial_z\Psi(z,\zeta,s) = -\Upsilon\t(z,s)\Psi(z,\zeta,s), \quad \Psi(\zeta,\zeta,s) = I. 
\end{equation}
With this, the solution of \eqref{statebvp} and \eqref{dode32m} is 
\begin{multline}
 \rho\t(z) = \rho_{ik,2}\t(1)E_2\t\Psi\t(z,1,\mu_i)\\ - \int_1^z h_{i(k-1)}\t(\zeta)E_1\t\Psi\t(z,\zeta,\mu_i)\d\zeta- p\t_{i(k)}E_1\t\Psi\t(z,1,\mu_i).  
\end{multline}
Inserting this in \eqref{dode22m} results in
\begin{equation}\label{solconddec}
 \rho_{ik,2}\t(1)E_2\t M(\mu_i) = r\t_{i(k-1)}
\end{equation}
with 
\begin{equation}\label{app:Mdef}
 M(s) = \Psi\t(0,1,s)E_1 + \int_0^1\Psi\t(\zeta,1,s) E_2\Lambda^{-1}(\zeta)\tilde{A}_0(\zeta)\d \zeta
\end{equation}
and 
\begin{align}\label{rbmdef}
&r\t_{i(k-1)} = \int_0^1 \Big(
h_{i(k-1)}\t(z)E_1\t\Psi\t(0,z,\mu_i)E_1\nonumber\\
& + \int_1^zh_{i(k-1)}\t(z)(\zeta)E_1\t \Psi\t(z,\zeta,\mu_i)E_2\d\zeta\,\Lambda^{-1}(z)\tilde{A}_0(z)\Big)\d z\nonumber\\
& - p\t_{i(k)}E_1\t\Psi\t(0,1,\mu_i)E_1\nonumber\\ 
& + \int_0^1p\t_{i(k)}E_1\t\Psi\t(z,1,\mu_i)E_2\Lambda^{-1}(z)\tilde{A}_0(z)\d z.
\end{align}
Hence, a solution of \eqref{firstIVP} exists, if
\begin{equation}\label{app:soldecouplssfeedsig2}
 \det E_2\t M(\mu_i) \neq 0, \quad i = 1,2,\ldots,r.
\end{equation}
It is shown in \cite{Deu17} that the backstepping target system \eqref{pdesc}--\eqref{bc2c} has a discrete point spectrum. Hence, the corresponding eigenvalue problem for the eigenvectors $\phi_i$ w.r.t. the eigenvalues $\tilde{\lambda}_i$ reads
\begin{subequations}\label{tfdet2}
	\begin{align}
	\phi_i''(z)  &= \Lambda^{-1}(z)\big((\tilde{\lambda}_i + \mu_c)\phi_i(z) + \tilde{A}_0(z)\phi_i(0)\big)\label{contrcasode2}\\
	\phi'_i(0) &= 0\label{inivarrho}\\
	\phi'_i(1) &= 0\label{rbvarrho}
	\end{align}
\end{subequations}
for $i \in \mathbb{N}$, in which \eqref{contrcasode2} is defined on $z \in (0,1)$. With the state $\varrho_i = \col{\varrho_{1,i},\varrho_{2,i}} = \col{\phi_i,\phi'_i}$ the ODE \eqref{contrcasode2} can be represented by
\begin{equation}\label{varrhodgl}
\varrho'_i(z) = \Upsilon(z,\tilde{\lambda}_i)\varrho(z) + E_2\Lambda^{-1}(z)\tilde{A}_0(z)\varrho_{1,i}(0),
\end{equation}
in which $\Upsilon(z,s)$ is given by \eqref{Adef}. Hence, the related fundamental matrix $\Phi(z,\zeta,s)$ is the solution of the IVP
\begin{equation}\label{Phidef}
\partial_z\Phi(z,\zeta,s) = \Upsilon(z,s)\Phi(z,\zeta,s), \quad \Phi(\zeta,\zeta,s) = I. 
\end{equation}
With this, the solution of the IVP \eqref{varrhodgl} and \eqref{inivarrho} can be represented by
\begin{equation}\label{ivpvarrhosol}
\varrho_{i}(z) = \bar{M}(z,\tilde{\lambda}_i)\varrho_{1,i}(0)
\end{equation}
where 
\begin{equation}\label{app:Mbardef}
 \bar{M}(z,s)
 	= \Phi(z,0,s)E_1 +  \int_0^z\Phi(z,\zeta,s)E_2\Lambda^{-1}(\zeta)\tilde{A}_0(\zeta)\d\zeta.
\end{equation}
By inserting this in \eqref{rbvarrho} one obtains the condition
\begin{equation}
 \varrho_{2,i}(1) =  E_2\t\bar{M}(1,\tilde{\lambda}_i)\varrho_{1,i}(0) = 0
\end{equation}
to determine $\varrho_{1,i}(0)$. Hence, nontrivial solutions $\varrho_{1,i}(0)$ and thus nonvanishing eigenvectors $\phi_i$ exist only if $\tilde{\lambda}_i$ is a solution of the \emph{characteristic equation}
\begin{equation}\label{app:chareq}
 \det (E_2\t\bar{M}(1,s)) = 0, \quad s \in \mathbb{C}.
\end{equation}
In order to relate this result to the solvability of \eqref{firstIVP}, a relationship is established between the fundamental matrices $\Psi$ in \eqref{Psidef} and $\Phi$ resulting from \eqref{Phidef}. For this, treat $\zeta$ as a parameter and consider $\d_z(\Psi\t\Phi) = \Psi_z\t\Phi + \Psi\t\Phi_z = -\Psi\t\Upsilon\Phi + \Psi\t\Upsilon\Phi = 0$ in view of \eqref{Psidef} and \eqref{Phidef}. Hence, $\Psi\t\Phi = \textit{const.}$ holds, which is the well-known \emph{reciprocity relation} between the fundamental matrices of the primal IVP \eqref{Phidef} and the related adjoint IVP \eqref{Psidef} (see, e.\:g., \cite[Ch. 4]{Ru96}). From this,  	
\begin{align}
\int_{\zeta}^{z}\d_{\bar{z}}(\Psi\t(\bar{z},\zeta,s)\Phi(\bar{z},\zeta,s))\d\bar{z} &= \Psi\t(z,\zeta,s)\Phi(z,\zeta,s) - I\nonumber\\
&= 0
\end{align}
is readily obtained in light of \eqref{Psidef} and \eqref{Phidef}. Using the latter result and the well-known property $(\Psi(z,\zeta,s))^{-1} = \Psi(\zeta,z,s)$ of a fundamental matrix yield the relationship
\begin{equation}\label{relfm}
\Phi(z,\zeta,s) = (\Psi\t(z,\zeta,s))^{-1} = \Psi\t(\zeta,z,s).
\end{equation}
By utilizing \eqref{relfm}, a comparison of \eqref{app:Mdef} and \eqref{app:Mbardef} shows that $\bar{M}(1,s) = M(s)$. In view of \eqref{app:soldecouplssfeedsig2} and \eqref{app:chareq} this leads to the condition $\sigma(F_{22}) \cap \sigma_c = \emptyset$ of Lemma \ref{lem:bvpsf}, while a piecewise $C^2$-solution of \eqref{firstIVP} follows from the fact that the elements of $H(z)$ in \eqref{dode12} are piecewise $C^1$. Then, by making use of the fact that the generalized eigenvectors $\varphi_{i(k)}$ are linearly independent, the solution
\begin{equation}
\Sigma_2(z) = \begin{bmatrix}
\varphi\t_{1(1)}\\
\vdots\\
\varphi\t_{1(l_1)}\\
\vdots\\
\varphi\t_{r(l_r)}
\end{bmatrix}^{-1}
\begin{bmatrix}
\sigma\t_{1(1)}(z)\\
\vdots\\
\sigma\t_{1(l_1)}(z)\\
\vdots\\
\sigma\t_{r(l_r)}(z)
\end{bmatrix}
\end{equation}
of \eqref{firstIVP} is obtained. The Sylvester equation \eqref{syleq} is uniquely solvable, if 
$\sigma(F_{22}) \cap \sigma(\tilde{F}_0) = \emptyset$ (see, e.\:g., \cite[Ch. 12.5, Th. 2]{La85}).
\hfill $\blacksquare$

\subsection{Proof of Theorem \ref{thm:sloop}}\label{app:{thm:sloop}}
In order to facilitate the stability analysis, the ODE-PDE-ODE cascade \eqref{drsc} is decoupled into an ODE and a PDE system. Towards this end, consider the change of coordinates 
\begin{equation}\label{app:B:firsttraf}
 \del\tilde{\varepsilon}(z,t) = \tilde{\varepsilon}(z,t) - 
 \Omega_1(z)\tilde{w}_1(t)
\end{equation}
with $\Omega_1(z) \in \mathbb{R}^{n \times n_1}$ to map \eqref{drsc} into the target system
\begin{subequations}\label{drsca}
	\begin{align}
	\dot{\tilde{w}}_1(t) &= \tilde{F}_1\tilde{w}_1(t) \label{plantode2ca}\\
	\partial_t\del\tilde{\varepsilon}(z,t) &= \Lambda(z)\partial_z^2\del\tilde{\varepsilon}(z,t) - \mu_c\del\tilde{\varepsilon}(z,t) - \tilde{A}_0(z)\del\tilde{\varepsilon}(0,t)\label{pdesca}\\
	\partial_z\del\tilde{\varepsilon}(0,t) &= 0\label{bc1ca}\\
	\partial_z\del\tilde{\varepsilon}(1,t) &= 0\label{bc2ca}\\
	\dot{w}_0(t) &= \tilde{F}_0w_0(t) \! + \! B_0\Omega_1(0)\tilde{w}_1(t) \! + \! B_0\del\tilde{\varepsilon}(0,t).\label{plantodeca}		
	\end{align}
\end{subequations}
Therein, the $\del\tilde{\varepsilon}$-system is decoupled from the $\tilde{w}_1$-system. For this, the matrix $\Omega_1(z)$ in \eqref{app:B:firsttraf} has to satisfy the BVP
\begin{subequations}\label{firstIVPa}
	\begin{align}
	\Lambda(z)\Omega''_1(z)  -  \mu_c\Omega_1(z) - \Omega_1(z)\tilde{F}_{1} &=  \tilde{A}_0(z)\Omega_1(0)\label{odedecw1a}\\
	\Omega'_1(0) &= 0\\
	\Omega'_1(1) &= D,
	\end{align}	
\end{subequations}   
in which \eqref{odedecw1a} is defined on $z \in (0,1)$. With the same reasoning as in Appendix \ref{app:lem:bvpsf}, it can be shown that \eqref{firstIVPa} has a unique piecewise $C^2$-solution if $\sigma(\tilde{F}_1) \cap \sigma_c = \emptyset$. Next, the $w_0$-system \eqref{plantodeca} is decoupled from the $\del\tilde{\varepsilon}$-system \eqref{pdesca}--\eqref{bc2ca} by making use of 
\begin{equation}\label{app:2trafsbew}
 \del w_0(t) = w_0(t) - \int_0^1\Omega_2(\zeta)\del\tilde{\varepsilon}(\zeta,t)\d\zeta
\end{equation}
with $\Omega_2(z) \in \mathbb{R}^{n_0 \times n}$. This results in
\begin{subequations}\label{drscaa}
	\begin{align}
    \begin{bmatrix}
     \dot{\tilde{w}}_1(t)\\
     \del \dot{w}_0(t)
    \end{bmatrix}
    &=
    \begin{bmatrix}
     \tilde{F}_1 & 0\\
     B_0\Omega_1(0)  & \tilde{F}_0
    \end{bmatrix}
    \begin{bmatrix}
     \tilde{w}_1(t)\\
     \del w_0(t)
    \end{bmatrix}\label{app:B:w0delw1sys}\\	
	\partial_t\del\tilde{\varepsilon}(z,t) &= \Lambda(z)\partial_z^2\del\tilde{\varepsilon}(z,t) - \mu_c\del\tilde{\varepsilon}(z,t) - \tilde{A}_0(z)\del\tilde{\varepsilon}(0,t)\label{pdescaa}\\
	\partial_z\del\tilde{\varepsilon}(0,t) &= 0\label{bc1caa}\\
	\partial_z\del\tilde{\varepsilon}(1,t) &= 0.\label{bc2caa}
	\end{align}
\end{subequations}
The matrix $\Omega_2(z)$ in \eqref{app:2trafsbew} has to be the solution of 
\begin{subequations}\label{firstIVPaaa}
	\begin{align}
	&(\Omega_2\Lambda)''(z)  -  \mu_c\Omega_2(z) - \tilde{F}_0\Omega_2(z) = 0, \quad z \in (0,1)\label{odedecw1aaa}\\
	&(\Omega_2\Lambda)'(0) =  \textstyle\int_0^1\Omega_2(\zeta)\tilde{A}_0(\zeta)\d \zeta  + B_0\\
	&(\Omega_2\Lambda)'(1) = 0.
	\end{align}	
\end{subequations}    
Since this BVP has the same structure as \eqref{firstIVP}, Appendix \ref{app:lem:bvpsf} implies that \eqref{drscaa} has a unique piecewise $C^2$-solution, if $\sigma(\tilde{F}_0) \cap \sigma_c = \emptyset$. 

Obviously, the $(\tilde{w}_1,\del w_0)$-system \eqref{app:B:w0delw1sys} is exponentially stable with the growth rate $\max(\alpha_{c,1},\alpha_{c,0})$, if $\tilde{F}_1$ and $\tilde{F}_0$ are Hurwitz matrices. In order to investigate the stability of the $\del\tilde{\varepsilon}$-system, introduce the state $\del\tilde{\varepsilon}(t) = \{\del\tilde{\varepsilon}(z,t), z \in [0,1]\}$ in the state space $X = (L_2(0,1))^n$ with the usual weighted inner product inducing the norm $\|\cdot\|_{L_2^n}$. Then, the PDE subsystem \eqref{pdescaa}--\eqref{bc2caa} can be represented by the abstract IVP 
\begin{equation}
 \del\dot{\tilde{\varepsilon}}(t) = \mathcal{A}_c\del\tilde{\varepsilon}(t), \quad t > 0,\; \del\tilde{\varepsilon}(0) \in D(\mathcal{A}_c) \subset X,
\end{equation}
with $\mathcal{A}_ch = \Lambda h'' - \mu_ch - \tilde{A}_0h(0)$ and
\begin{equation}
 D(\mathcal{A}_c) = \{h \in (H^2(0,1))^n\;|\;h'(0) = h'(1) = 0\}.
\end{equation}
It is shown in \cite{Deu18} that $\mathcal{A}_c$ is the generator of an analytic $C_0$-semigroup, that is exponentially stable with the growth rate $\alpha_{\tilde{\varepsilon}}$. Consequently,
\begin{equation}\label{thexpstabsfeeda}
\|\del\tilde{\varepsilon}(t)\|_{L_2^n} \leq M\e^{(\alpha_{\tilde{\varepsilon}} + c)t}	
\|\del\tilde{\varepsilon}(0)\|_{L_2^n},\quad  t \geq 0,
\end{equation}
holds for all $\del\tilde{\varepsilon}(0) \in D(\mathcal{A}_c)$, an $M \geq 1$ and any $c > 0$ such that $\alpha_{\tilde{\varepsilon}} + c < 0$. With these results, it is straightforward to verify that \eqref{drscaa} is exponentially stable in the norm $\|\cdot\|$ with the decay rate $\alpha_c$. Then, by making use of all involved transformations and their bounded invertibility, standard arguments yield the stability result of Theorem \ref{thm:sloop}. \hfill $\blacksquare$

\subsection{Proof of Theorem \ref{thm:obs}}\label{app:theo:obs}
Similar to the proof of Theorem \ref{thm:sloop}, the ODE-PDE-ODE cascade \eqref{obsef} is decoupled into an ODE and a PDE. For this, the $\varepsilon_x$-system in  \eqref{obsef} is decoupled from the $\tilde{e}_1$-system. Introducing the change of coordinates 
\begin{equation}\label{app:B:firsttrafb}
\del\varepsilon_x(z,t) = \varepsilon_x(z,t) - 
\Theta_1(z)\tilde{e}_1(t)
\end{equation}
with $\Theta_1(z) \in \mathbb{R}^{n \times n_1}$, the target system 
\begin{subequations}\label{obsefa}
	\begin{align}
	\dot{\tilde{e}}_1(t) &= \tilde{\bar{F}}_1\tilde{e}_1(t)\label{plantode2oefa}\\
	\partial_t\del\varepsilon_x(z,t) &= \Lambda(z)\partial^2_z\del\varepsilon_x(z,t) - \mu_o\del\varepsilon_x(z,t)\label{xeqoefa}\\
	\partial_z\del\varepsilon_x(0,t) &= 0\\
	\partial_z\del\varepsilon_{x}(1,t) &= -\textstyle\int_0^1\tilde{\bar{A}}_0(\zeta)\del\varepsilon_x(\zeta,t)\d\zeta\label{uabcoefa}\\
	\dot{\tilde{e}}_0(t) &= \tilde{\bar{F}}_0\tilde{e}_0(t) \! + \! \bar{D}\Theta_1(0)\tilde{e}_1(t) \! + \! \bar{D}\del\varepsilon_{x}(0,t)\label{plantodeoefa} 
	\end{align}
\end{subequations}
can be considered. Then, the matrix $\Theta_1(z)$ in \eqref{app:B:firsttrafb} has to be the solution of the BVP
\begin{subequations}\label{firstIVPao}
	\begin{align}
	&\Lambda(z)\Theta''_1(z)  -  \mu_o\Theta_1(z) - \Theta_1(z)\tilde{\bar{F}}_{1} = 0, \quad z \in (0,1)\label{odedecw1ao}\\
	&\Theta'_1(0) = 0\\
	&\Theta'_1(1) = C_1 - \textstyle\int_0^1\tilde{\bar{A}}_0(\zeta)\Theta_1(\zeta)\d\zeta.
	\end{align}	
\end{subequations}   
By making use of the same procedure as in Appendix \ref{app:lem:bvpsf}, it can be verified that \eqref{firstIVPao} has a unique piecewise $C^2$-solution, if $\sigma(\tilde{\bar{F}}_1) \cap \sigma_o = \emptyset$. In the next step, the $\tilde{e}_0$-system \eqref{plantodeoefa} is decoupled from the $\del\varepsilon_x$-system \eqref{xeqoefa}--\eqref{uabcoefa}. This can be achieved with the transformation
\begin{equation}\label{app:2trafsbew0}
\del\tilde{e}_0(t) = \tilde{e}_0(t) - \int_0^1\Theta_2(\zeta)\del\varepsilon_x(\zeta,t)\d\zeta
\end{equation}
where $\Theta_2(z) \in \mathbb{R}^{n_0 \times n}$. As a results, one obtains
\begin{subequations}\label{drscaao}
	\begin{align}
	\begin{bmatrix}
	\dot{\tilde{e}}_1(t)\\
	\del \dot{\tilde{e}}_0(t)
	\end{bmatrix}
	&=
	\begin{bmatrix}
	\tilde{\bar{F}}_1 & 0\\
	\bar{D}\Theta_1(0)  & \tilde{\bar{F}}_0
	\end{bmatrix}
	\begin{bmatrix}
	\tilde{e}_1(t)\\
	\del \tilde{e}_0(t)
	\end{bmatrix}\label{app:B:w0delw1syso}\\	
	\partial_t\del\varepsilon_x(z,t) &= \Lambda(z)\partial_z^2\del\varepsilon_x(z,t) - \mu_o\del\varepsilon_x(z,t)\label{pdescaao}\\
	\partial_z\del\varepsilon_x(0,t) &= 0\label{bc1caao}\\
	\partial_z\del\varepsilon_x(1,t) &= -\textstyle\int_0^1\tilde{\bar{A}}_0(\zeta)\del\varepsilon_x(\zeta,t)\d\zeta.\label{bc2caao}
	\end{align}
\end{subequations}
For this, the matrix $\Theta_2(z)$ in \eqref{app:2trafsbew0} has to solve
\begin{subequations}\label{firstIVPaaa2}
	\begin{align}
	&(\Theta_2\Lambda)''(z)  \! - \! \mu_o\Theta_2(z) \! - \! \tilde{\bar{F}}_0\Theta_2(z) = (\Theta_2\Lambda)(1)\tilde{\bar{A}}_0(z)\label{odedecw1aaa2}\\
	&(\Theta_2\Lambda)'(0) = \bar{D}\\
	&(\Theta_2\Lambda)'(1) = 0,
	\end{align}	
\end{subequations}    
in which \eqref{odedecw1aaa2} is defined on $z \in (0,1)$. This BVP is of the same form as \eqref{firstIVP}. Hence, with the results of Appendix \ref{app:lem:bvpsf} one finds that a unique piecewise $C^2$-solution of \eqref{firstIVPaaa2} exists, if $\sigma(\tilde{\bar{F}}_0) \cap \sigma_o = \emptyset$. 

Assume that $\tilde{\bar{F}}_1$ and $\tilde{\bar{F}}_0$ are Hurwitz matrices. Then, the  $(\tilde{e}_1,\del \tilde{e}_0)$-system \eqref{app:B:w0delw1syso} is exponentially stable with the growth rate $\max(\alpha_{o,1},\alpha_{o,0})$. The stability of the $\del\varepsilon_x$-system is investigated on the basis of its state space representation. To this end, introduce the state $\del\varepsilon_x(t) = \{\del\varepsilon_x(z,t), z \in [0,1]\}$ in the state space $X = (L_2(0,1))^n$ with the usual weighted inner product inducing the norm $\|\cdot\|_{L_2^n}$. Then, the PDE subsystem \eqref{pdescaao}--\eqref{bc2caao} leads to the abstract IVP 
\begin{equation}
\del\dot{\varepsilon}_x(t) = \mathcal{A}_o\del\varepsilon_x(t), \quad t > 0,\; \del\varepsilon_x(0) \in D(\mathcal{A}_o) \subset X,
\end{equation}
with $\mathcal{A}_oh = \Lambda h'' - \mu_oh$ and
\begin{multline}
 D(\mathcal{A}_o) = \{h \in (H^2(0,1))^n\;|\; h'(0) = 0,\\ h'(1) = -\textstyle\int_0^1\tilde{\bar{A}}_0(\zeta)h(\zeta)\d\zeta\}.
\end{multline}
It is verified in \cite{Deu18} that $\mathcal{A}_o$ is the generator of an analytic $C_0$-semigroup, that is exponentially stable with the growth rate $\alpha_{\varepsilon_x}$. This leads to
\begin{equation}\label{thexpstabsfeeda2}
\|\del\varepsilon_x(t)\|_{L_2^n} \leq M\e^{(\alpha_{\varepsilon_x} + c)t}	
\|\del\varepsilon_x(0)\|_{L_2^n},\quad  t \geq 0,
\end{equation}
satisfied for all $\del\varepsilon_x(0) \in D(\mathcal{A}_o)$, an $M \geq 1$ and any $c > 0$ such that $\alpha_{\varepsilon_x} + c < 0$. By making use of these results, the exponential stability of  \eqref{drscaao} in the norm $\|\cdot\|$ with the decay rate $\alpha_o$ can be shown. Then, by taking all involved transformations and their bounded invertibility into account, the stability result of Theorem \ref{thm:obs} is deducible from standard arguments. \hfill $\blacksquare$

\subsection{Proof of Theorem \ref{thm:cloop}}\label{app:thm:cloop}
Represent the closed-loop system by the observer error dynamics \eqref{drscaao} and the observer \eqref{obs} with 
\begin{equation}
 \Delta(t) = y(t) - C\hat{w}_0(t) = C\bar{T}\tilde{e}_0(t)
\end{equation}
in view of \eqref{meas}, \eqref{inno} and \eqref{ccdecouplo}. Apply the sequence of transformations shown in Figure \ref{fig:tseq} to the observer and use the feedback \eqref{sfeedest}. Subsequently, use the transformations of Appendix \ref{app:{thm:sloop}} for the resulting system, in order to decouple the PDE subsystem and the ODE subsystem at $z = 0$. This results in
\begin{subequations}\label{drscaap}
	\begin{align}
	\begin{bmatrix}
	\dot{\tilde{w}}^{\mathit{obs}}_1(t)\\
	\del \dot{w}^{\mathit{obs}}_0(t)
	\end{bmatrix}
	&=
	\begin{bmatrix}
	\tilde{F}_1 & 0\\
	B_0\Omega_1(0)  & \tilde{F}_0
	\end{bmatrix}
	\begin{bmatrix}
	\tilde{w}^{\mathit{obs}}_1(t)\\
	\del w^{\mathit{obs}}_0(t)
	\end{bmatrix}\nonumber\\
	& \quad  + G_1\tilde{e}_0(t) + H_1\dot{\tilde{e}}_0(t)\label{app:B:w0delw1sysp}\\	
	\partial_t\del\tilde{\varepsilon}^{\mathit{obs}}(z,t) &= \Lambda(z)\partial_z^2\del\tilde{\varepsilon}^{\mathit{obs}}(z,t) - \mu_c\del\tilde{\varepsilon}^{\mathit{obs}}(z,t)\nonumber\\
	& \quad  - \tilde{A}_0(z)\del\tilde{\varepsilon}^{\mathit{obs}}(0,t) 
	  + G_2(z)\tilde{e}_0(t)\nonumber\\
	& \quad  +  G_3(z)\dot{\tilde{e}}_0(t)\label{pdescaap}\\
	\partial_z\del\tilde{\varepsilon}^{\mathit{obs}}(0,t) &= L^0_0C\bar{T}\tilde{e}_0(t) + L^1_0C\bar{T}\dot{\tilde{e}}_0(t) \label{bc1caap}\\
	\partial_z\del\tilde{\varepsilon}^{\mathit{obs}}(1,t) &= L_1C\bar{T}\tilde{e}_0(t)\label{bc2caap}
	\end{align}
\end{subequations}
for some matrices $G_1$ and $G_2(z)$. Solving \eqref{app:2trafsbew0} for $\tilde{e}_0$ gives
\begin{equation}\label{app:2trafsbew02}
 \tilde{e}_0(t) = \;\del\tilde{e}_0(t) + \int_0^1\Theta_2(\zeta)\del\varepsilon_x(\zeta,t)\d\zeta.
\end{equation}
Hence, the closed-loop system can be represented by a cascade of two exponentially stable systems \eqref{drscaao}, \eqref{drscaap}  and \eqref{app:2trafsbew02} in the new coordinates. It can be shown that the PDE subsystems \eqref{pdescaao}--\eqref{bc2caao} and \eqref{pdescaap}--\eqref{bc2caap} have the same structure as the corresponding result in \cite{Deu18}.  Hence, the closed-loop system can be represented by the abstract IVP
\begin{equation}\label{app:clivp}
 \dot{\tilde{x}}_{\mathit{cl}}(t) = \mathcal{A}_{\mathit{cl}}\tilde{x}_{\mathit{cl}}(t), \quad t > 0, \tilde{x}_{\mathit{cl}}(0) \in D(\mathcal{A}_{\mathit{cl}}) \subset X_{\mathit{cl}}
\end{equation}
in the state space $X_{\mathit{cl}} = \mathbb{C}^{n_0} \oplus \mathbb{C}^{n_1} \oplus (H^2(0,1))^n \oplus \mathbb{C}^{n_0} \oplus \mathbb{C}^{n_1} \oplus (H^2(0,1))^n$ with the usual weighted inner product inducing the norm $\|\cdot\|_{\mathit{cl}}$ and the closed-loop state $\tilde{x}_{\mathit{cl}}(t) = \operatorname{col}(\del\tilde{e}_0(t),\tilde{e}_1(t),\del\varepsilon_x(t),\del w^{\mathit{obs}}_0(t),\tilde{w}^{\mathit{obs}}_1(t),\del\tilde{\varepsilon}^{\mathit{obs}}(t))$. By making use of the corresponding results in \cite{Deu18}, it can be verified that  $\mathcal{A}_{\mathit{cl}}$ is the generator of an exponentially stable and analytic $C_0$-semigroup with the growth rate $\alpha_{\mathit{cl}}$. Then, after going through the corresponding chain of boundedly invertible transformations, the stability result \eqref{thexpstabsclfeed} in the original coordinates is obtained. \hfill $\blacksquare$


\section*{Acknowledgment}
The work of the second author has been supported by the Pro$^2$Future competence center in the framework of the Austrian COMET-K1 program under contract no.\ 854184.

\bibliographystyle{IEEEtranS}
\bibliography{IEEEabrv,mybib}

\vspace*{-2\baselineskip}
\begin{IEEEbiography}[{\includegraphics[width=1in,height=1.25in,clip,keepaspectratio]{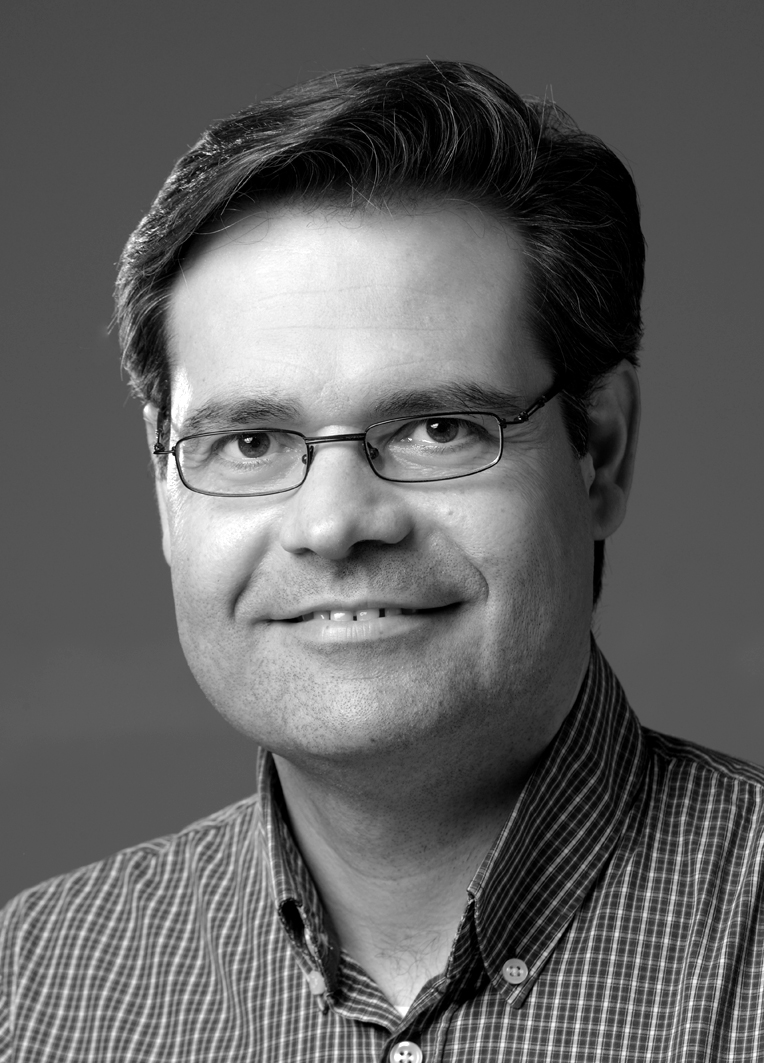}}]{Joachim Deutscher} (M '18)
received the Dipl.-Ing. (FH) degree in electrical engineering from the University of Applied Sciences W\"urzburg-Schweinfurt-Aschaffenburg, Germany, in 1996, the Dipl.-Ing. Univ. degree in electrical engineering, the Dr.-Ing. and the Dr.-Ing. habil. degrees both in automatic control from the Friedrich-Alexander Universit\"at Erlangen-N\"urnberg (FAU), Germany, in 1999, 2003 and 2010, respectively.

From 2003--2010 he was a Senior Researcher at the Chair of Automatic Control (FAU), in 2011 he was appointed Associate Professor and in 2017 Professor at the same university. Since April 2020 he is a Full Professor at the Institute of Measurement, Control and Microtechnology, Ulm University.

His research interests include control of distributed-parameter systems and control theory for nonlinear lumped-parameter systems with applications in mechatronic systems and robotics. Dr. Deutscher has co-authored a book on state feedback control for linear lumped-parameter systems: Design of Observer-Based Compensators (Springer, 2009) and is author of the book: State Feedback Control of Distributed-Parameter Systems (in German) (Springer, 2012). At present he serves as Associate Editor for Automatica. 
\end{IEEEbiography}

\begin{IEEEbiography}[{\includegraphics[width=1in,height=1.25in,clip,keepaspectratio]{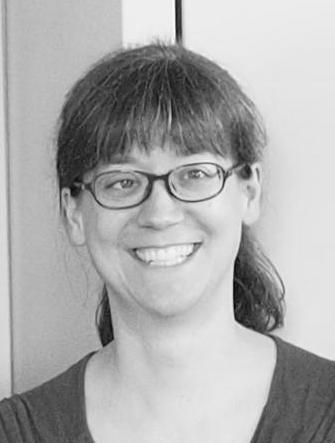}}]{Nicole Gehring} received the Dipl.-Ing. degree in electrical engineering from Dresden University of Technology, Germany, in 2007. Following a two-year stint of industrial work in the control of power plants and motor vehicles, in 2015, she finished her doctoral thesis at Saarland University, Germany, and received the Dr.-Ing. degree.

As a Postdoc, she stayed with the Technical University of Munich, Germany, for two year. Currently, she is with Johannes Kepler University Linz, Austria.
	
Her research mainly focuses on linear distributed-parameter systems and linear time-delay systems, especially in the context of control and observer design, as well as parameter identification.	
\end{IEEEbiography}
\vfill	
\end{document}